\documentclass[superscriptaddress,amssymb,amsmath,nofootinbib,aps,showkeys,floatfix,prd,a4paper,twocolumn]{revtex4-1}

\usepackage{graphicx}
\usepackage{hyperref}
\usepackage{xcolor}
\hypersetup{colorlinks=true, linkcolor=blue, citecolor=blue}
\usepackage[toc,page]{appendix}

\usepackage{placeins}
\usepackage{multirow}

\newcommand{\+}{\hspace{-1mm}+\hspace{-1mm}}

\begin{document}


\title{Quarkonium interactions with (hot) hadronic matter}

\author{Luciano M. Abreu}
\email{luciano.abreu@ufba.br}
\affiliation{Instituto de F\'isica, Universidade Federal da Bahia, Campus Universit\'ario de Ondina, Salvador, Bahia, 40170-115, Brazil}

\author{Hildeson P. L. Vieira}
\email{hilde_son@hotmail.com}
\affiliation{Instituto de F\'isica, Universidade Federal da Bahia, Campus Universit\'ario de Ondina, Salvador, Bahia, 40170-115, Brazil}

\begin{abstract}

In this work, we present an updated study about the interactions of quarkonia with surrounding hadronic medium. The meson-meson interactions are described with a chiral effective Lagrangian within the framework of unitarized coupled channel amplitudes. In particular, we extend a previous work performed in the charmonium sector by calculating the cross-sections for $\Upsilon$ scattering by light pseudoscalar mesons ($\pi, K, \eta$) and vector mesons ($\rho, K^\ast, \omega$). We evaluate the relevant channels and compare the results with existing literature. 
The analysis is completed by including the finite-temperature effects in the unitarized  scattering amplitudes for both quarkonium and bottomonium sectors.

\end{abstract}

\maketitle

\section{Introduction} 
\label{Introduction} 

In the latest decades we have observed immense progresses in the understanding of heavy-ion collisions. In this scenario, measurements have shown that heavy-flavored hadrons play a major role in the characterization of  the evolution of partonic matter. This comes from the fact that heavy quarks are yielded by hard gluons in the initial stages of collision, with a quark-gluon plasma (QGP) being formed surrounding them, and therefore undergoing the whole evolution of the system. As the QGP expands and cools down, the system hadronizes and is not hot enough to excite heavy-quark pairs, in contrast to light hadrons which can be produced in the medium at later stages.

Based on the prediction in the 1980's that in hot and dense mediums, like the QGP, charmonia (bound states of charm $c$ and anti-charm $\bar{c}$ quarks) are suppressed because of screening of the $c-\bar{c}$ interaction, engendered by the high density of color charges~\cite{Matsui:1986dk,RAPP2010209,Braun-Munzinger}, the $J/\psi$ has became a valuable probe of properties of QGP. 
Indeed, a number of Collaborations at RHIC, SPS and LHC have searched and detected evidences of  $J/\psi$ suppression~\cite{Adams:2005dq,Alessandro:2004ap,Abelev:2013ila,Adam:2016rdg}. Moreover, the suppression effect has also been observed in the bottomoium sector, in particular in the sequential suppression of the $\Upsilon (1S)$, $\Upsilon (2S)$ and $\Upsilon (3S)$ states in Pb–Pb collisions at LHC~\cite{Chatrchyan:2012lxa}. 

However, the analysis of data on $J/\psi$ production in the LHC era has revealed that the QGP dynamics is richer and more involved than expected. At low transverse momentum ($p_T$), the $J/\psi$ suppression is smaller at LHC energies when compared with RHIC, indicating a  regeneration mechanism due to larger total charm cross section at LHC; on the other hand, at high $p_T$ the dissociation increases for higher collision energies, suggesting that the  $J/\psi$ production is not too sensitive to recombination and other effects~\cite{Abelev:2013ila,Adam:2016rdg,Zha:2017xsm}. 

In parallel, other proposals have been put forward to describe the heavy quarkonia properties in heavy-ion collisions. One of them, which is the subject of the present work, is their interaction with hadronic medium. Noticing that the quarkonia are created at the 
start of collision, those that have gotten through the QGP phase can experience inelastic interactions with other particles composing the hadronic matter. Thus, this mechanism  may be relevant as a part of the explanation of the quarkonium multipliplicity evolution.

A significant quantity of investigations, by means of different approaches (e.g. effective hadron Lagrangians, constituent quark-model framework and others), has been devoted to provide information on the  quarkonium  interactions with light mesons in a hadron gas, being most of them involving the $J/\psi$
~\cite{Wong:1999zb,Wong:2001td,PhysRevC.58.2994,PhysRevC.61.031902,Braun-Munzinger2000,PhysRevC.62.034903,PhysRevC.63.065201,PhysRevC.63.034901,PhysRevC.68.014903,Oh:2002vg,PhysRevC.68.035208,Maiani:2004py,Maiani:2004qj,DURAES200397,PhysRevC.70.055203,PhysRevC.72.024902,PhysRevD.72.034002,Capella2008,CASSING20011,doi:10.1142/S0218301308010507,PhysRevC.85.064904,MITRA201675,Liu2016,PhysRevC.96.045201,PhysRevC.97.044902,Abreu:2018mji}. In particular, in Ref.~\cite{Abreu:2018mji} the cross sections for a larger number of channels of the $J/\psi $-light meson scatterings with respect to precedent studies have been estimated within the framework of unitarized coupled channel amplitudes, with reactions carrying charmed mesons or charmonia in final state yielding the most relevant contributions.

But for the case of $\Upsilon$,  in opposition to the situation of $J/\psi$ mentioned above, the amount of similar analyses are scarce. As far as we are aware, Refs.~\cite{Lin:2000ke,Abreu:2018mnc} have calculated cross sections of processes involving $\Upsilon $ absorption by $ \pi, \rho$ mesons in a hot hadron gas. They make use of distinct  effective hadronic Lagrangians, form-factors and cutoffs at the interaction vertices. They found that the values of cross sections $\sigma_{ \Upsilon \pi}$  and $\sigma_{ \Upsilon \rho}$   are deeply influenced by the these choices of form factors, cutoffs and coupling constants. Besides, they found contrasting results: while  
\cite{Lin:2000ke} concludes that the absorption of $\Upsilon$  by comoving hadrons is 
negligible due to their small thermal averages, in \cite{Abreu:2018mnc} the anomalous interactions are taken into account and the $\Upsilon$ abundance may suffer a sizable reduction in hadronic phase.

Thus, by virtue of the theoretical and experimental progress on $\Upsilon$ phenomenology, we intend to contribute to this issue in another perspective,  with the extension of the calculation performed in Ref.~\cite{Abreu:2018mji} to the bottomonia sector. 
In the present work we estimate the cross-sections of $\Upsilon$ with light pseudoscalar mesons ($\pi, K, \eta$) and light vector mesons ($\rho, K^\ast, \omega$), within the framework of unitarized coupled channel amplitudes~\cite{Abreu:2018mji,Roca:2005nm,Gamermann:2006nm,Gamermann:2007fi,Dias:2014pva,Molina:2008jw,Abreu2011,Abreu2013a,PhysRevC.96.045201}. We analyze the magnitude of unitarized cross sections for the different channels, and compare our results with others mentioned above.  The investigation is completed by including the finite-temperature effects in the unitarized scattering amplitudes for both quarkonium and bottomonium sectors, using a simplified version of imaginary-time formalism and the thermal masses~\cite{Gao:2019idb}.

We organize this work as follows. In Section II is presented an overview of the effective $SU(4)$ model and the unitarized coupled channel amplitudes.  Results will be presented in Section III. Finite-temperature effects are discussed in Section IV. We summarize the results and conclusions in Section V. 

\section{The model}
\label{formalism} 

We start by describing briefly the formalism employed to calculate and analyze the cross sections
for the $Q - X$ interactions, where $Q$ denotes a quarkonium and $X$ a pseudoscalar or vector meson. It will be in the context of the chiral $SU(4)$ model, which has been successfully applied in different situations~\cite{Gamermann:2006nm,Gamermann:2007fi,Roca:2005nm,PhysRevC.96.045201,Abreu:2018mji}. The cases of scatterings involving charmed mesons are treated in Ref.~\cite{Abreu:2018mji}, so here we extend this approach to the bottom sector. For a detailed discussion we refer the reader to these mentioned references. Concentrating on the fundamental features, the effective Lagrangian is given by
 \begin{equation}
\mathcal{L}_{\text{int}} = -\frac{1}{4f^2} \text{Tr}\left(J^\mu \mathcal{J}_\mu \right) -\frac{1}{4f^2} \text{Tr}\left(\mathcal{J}^\mu \mathcal{J}_\mu \right),
\label{Lagr}
\end{equation}
where $f$ is the meson decay constant; the factor $f^2$ must be replaced by $\sqrt{f_{\pi}}$ for each light meson line in the corresponding vertex and $\sqrt{f_B}$ for each heavy one; $Tr (\cdots)$ represents the trace over the flavor indices; 
 $J^\mu = [P,\partial^\mu P]$ and $\mathcal{J}^\mu = [V^\nu, \partial^\mu V_\nu]$ are the pseudoscalar and vector currents, respectively, with $P$ and $V$ being $4 \times 4$ matrices carrying 15-plets of pseudoscalar and vector fields extended to the bottom sector as used in Ref.~\cite{Dias:2014pva}. But here we use an unmixed representation, that is 
\begin{eqnarray}
&  &  P = 
 \sum_{i=1}^{15} \frac{\varphi_i}{\sqrt{2}} \lambda_i=  \nonumber \\
& & \begin{pmatrix}
\frac{\pi^0}{\sqrt{2}}\+\frac{\eta}{\sqrt{6}} \+ \frac{\eta_b}{\sqrt{12}} & \pi^+ & K^+ & \bar B^0\\
\pi^- & -\frac{\pi^0}{\sqrt{2}}\+\frac{\eta}{\sqrt{6}}\+\frac{\eta_b}{\sqrt{12}} & K^0 & B^-\\
K^- & \bar K^0 & -2\frac{\eta}{\sqrt{6}}\+\frac{\eta_b}{\sqrt{12}} & B_s\\
B^0 & B^+ & B_s & -\frac{\sqrt{3}}{2} \eta_b
\end{pmatrix}; \,\,\,
\nonumber \\
& & V_\mu  =   
\sum_{i=1}^{15} \frac{v_{\nu i}}{\sqrt{2}} \lambda_i= \nonumber \\
& & \begin{pmatrix}
\frac{\rho^0}{\sqrt{2}}\+\frac{\omega}{\sqrt{6}}\+\frac{\Upsilon}{\sqrt{12}} & \rho^+ & K^{*+} & \bar B^{*0}\\
\rho^- & -\frac{\rho^0}{\sqrt{2}}\+\frac{\omega}{\sqrt{6}}\+\frac{\Upsilon}{\sqrt{12}} & K^{*0} & B^{*-}\\
K^{*-} & \bar K^{*0} & -2\frac{\omega}{\sqrt{6}}\+\frac{\Upsilon}{\sqrt{12}} & B_s^{*}\\
B^{*0} & B^{*+} & B_s^{*} & -\frac{\sqrt{3}}{2} \Upsilon
\end{pmatrix}_\mu; \nonumber \\
\label{PV}
\end{eqnarray}
$\lambda_i$ being the generalized Gell-Mann matrices for $SU(4)$.

Keeping in mind that the exact $SU(4)$ symmetry in effective Lagrangian (\ref{Lagr}) is not reproduced in nature, it is properly broken here in accordance with previous works~\cite{Gamermann:2006nm,Gamermann:2007fi,Roca:2005nm,PhysRevC.96.045201,Abreu:2018mji}, outlined below. Invoking the vector-meson dominance ansatz, the interactions in this Lagrangian are interpreted in terms of a vector-meson exchange between two mesons. The $SU(4)$ symmetry supposes equal masses for all virtual vector-meson exchanged, and its breaking is implemented by the suppression of the terms in Eq.~(\ref{Lagr}) which induce a heavy vector-meson exchanged. In this sense, its propagator is approximated to be proportional to the inverse of its squared mass, $(1/m_V)^2$ (in the limit $q^2 \ll m_V^2$; $q$ is the transferred momentum). Thus, the suppressed terms 
are assumed to be of the order $(m_L / m_H)^2$, with $m_L$ and $m_H$ being the relevant scales of the light and heavy vector mesons masses, respectively.
Looking specifically the contributions in $\mathcal{L}_{\text{int}}$, those in which the two currents have bottom quantum number are connected through the exchange of a bottomed vector meson, and therefore are suppressed by introducing their corrected version with the multiplicative factor $\gamma = (m_L / m_H)^2$. 
In the case of the terms in $\mathcal{L}_{\text{int}}$ which allow the exchange of a heavy hidden bottom meson, where the $\rho$  and $\omega$ also take part: since the contributions involving light and heavy mesons connecting the currents have the respective weights $ -1/3$ and $4/3$ (they have been computed properly in Refs.~\cite{Gamermann:2006nm,Gamermann:2007fi}),  the suppression is performed inserting the factor $\psi = -1/3 + 4/3 (m_L/ m_H ^{\prime})^2$.  
The parameters $m_L$, $m_H$ and $m_H ^{\prime}$ are chosen appropriately, being related to the masses of  light vector-meson (e.g. the $\rho$ meson and $K^*$), bottomed vector-meson (like the $B^*$, $B_s ^*$ )  and bottomonium ($\Upsilon$), respectively.

We are interested in the scattering amplitudes of the $ \Upsilon X$ absorption processes: 
\begin{eqnarray}
(1) \;\; \Upsilon (p_1) P (p_2) & \rightarrow &  V (p_3) P (p_4) , \nonumber \\
(2) \;\; \Upsilon (p_1) V (p_2) & \rightarrow &  P (p_3) P (p_4) , \nonumber \\
(3) \;\;\Upsilon (p_1) V (p_2) & \rightarrow &  V (p_3) V (p_4) ,
\label{proc1}
\end{eqnarray}
where $p_j$  denotes the momentum of particle $j$, with particles 1 and 2 standing for initial state mesons, and particles 3 and 4 for final state mesons. Then, with the use of Eq.~(\ref{Lagr}), the invariant amplitudes for processes of type $ V P\rightarrow V P $ in Eq.~(\ref{proc1}) can be written as 
\begin{eqnarray}
\mathcal{M} _{1; i j}  (s, t, u )= \frac{\xi _{i j}  }{2f^2} (s-u)  \varepsilon _1 \cdot\varepsilon _3 ^{\ast} ,
\label{Eq:CasoVPVP}
\end{eqnarray}
while for processes $V V \rightarrow P P$ they are
\begin{eqnarray}
\mathcal{M} _{2; i j}  (s, t, u )= \frac{ \chi _{i j}  }{2f^2} (t-u)  \varepsilon _1 \cdot\varepsilon _2 ,
\label{Eq:CasoVVPP}
\end{eqnarray}
and for processes $V V \rightarrow V V$ we have, 
\begin{eqnarray}
\mathcal{M} _{3; i j}   (s, t, u ) & = & 
\frac{\zeta _{ i j } ^{(s)} }{2f^2} (t-u) \varepsilon _1 \cdot\varepsilon _2 \varepsilon ^{\ast} _3 \cdot\varepsilon _4 ^{\ast} \nonumber \\
& & + \frac{ \zeta _{ i j } ^{(t)} }{2f^2} (s-u)  \varepsilon_1\cdot\varepsilon _3 ^{\ast}  \varepsilon _2 \cdot\varepsilon _4 ^{\ast} \nonumber \\
& & + \frac{\zeta _{ i j } ^{(u)} }{2f^2} (s-u)  \varepsilon _1 \cdot\varepsilon _4 ^{\ast}  \varepsilon _2 \cdot\varepsilon _3 ^{\ast}, 
\label{Eq:CasoVVVV}
\end{eqnarray}
where the labels $i$ and $j$ refer to the initial
and final channels; $s, t$ and $u$ to the Mandelstam variables; $\varepsilon _a$ to the polarization vector related to the respective vector particle $a$.
The coefficients $\xi _{i j}, \chi _{i j } $ and $\zeta _{i j}$ will depend on the initial and final channels of each process, and are given in an isospin basis in Appendix~A of Ref.~\cite{Abreu:2018mji}, with the proper replacement of the charmed mesons by bottomed mesons.

We assume that these processes have their quantum numbers for the incoming and outcoming meson pairs conserved; namely, they are $I^G(J^{PC})$, charm ($C$) and strangeness ($S$). For the $s$-wave reactions of the channels involving pairs of vector mesons in Eq.~(\ref{Eq:CasoVVVV}), we decompose the polarization vectors of each incoming/outgoing pair $VV$ via spin-projectors~\cite{Roca:2005nm,Abreu:2018mji,Molina:2008jw}, which allow us to write a generic amplitude $\mathcal{A}$ decomposed as 
\begin{eqnarray}
\mathcal{A}^{(S=0)} & \equiv & \mathcal{P}_{a b} ^{(S=0) i i }\mathcal{P}_{c d} ^{(S=0) j j }, \nonumber \\
\mathcal{A}^{(S=1)}  & \equiv & \mathcal{P}_{a b } ^{(S=1)i j }\mathcal{P}_{c d }^{(S=1) i j } ,\nonumber \\
\mathcal{A}^{(S=2)}  & \equiv & \mathcal{P}_{a b} ^{(S=2) i j } \mathcal{P}_{c d} ^{(S=2) i j } ,
\label{ampA5}
\end{eqnarray}
where $\mathcal{P} ^{(S=0) }, \mathcal{P} ^{(S=1)}$ and $\mathcal{P} ^{(S=2)}$  are the spin-projectors associated to the scalar ($S=0$), antisymmetric tensor ($S=1$) and symmetric tensor ($S=2$) representations, respectively; the indices $i,j$ denote the component of polarization vector, and $ab(cd)$ labels the incoming(outgoing) pair of mesons.

Concerning the control mechanism of the energy-dependence of cross sections, previous works~\cite{Lin:2000ke,Abreu:2018mnc} obtained the correct behavior of the amplitudes at high energies for some reactions to be considered by making use of form factors with different functional forms and cutoff values without a more fundamental justification. In the present study we adopt a different way, based on the method of unitarized coupled channel amplitudes. This approach has been satisfactorily employed for the description hadronic resonances and meson-meson scatterings~\cite{Weinstein:1990gu,Janssen:1994wn,Oller:1997ti,Oller:1998hw,Roca:2005nm,Gamermann:2007fi,Abreu2011,Abreu2013a,PhysRevC.96.045201}.
Accordingly, the matrix representing the unitarized coupled channel transitions are obtained through a Bethe-Salpeter equation that can be written as 
\begin{equation}
\mathcal{T} (s)= \frac{V (s)}{1 + V(s) G(s)},
\label{Tmatrix}
\end{equation} 
whose kernel $V(s)$ is the s-wave projected scattering amplitude, 
\begin{equation}
V_{r; i j } (s) = \frac{1}{2} \int_{-1}^{1} d(\cos \theta) \mathcal{M} _{r; i j} \left( s, t(s,\cos \theta), u(s,\cos \theta)\right),
\label{Vamp}
\end{equation} 
with $r=1,2,3$ associating $\mathcal{M} _{r; i j}$ to a given process in Eqs.(\ref{Eq:CasoVPVP}), (\ref{Eq:CasoVVPP}) or (\ref{Eq:CasoVVVV}); $G (s)$ is the two-meson loop integral, which in the case of two pseudoscalars mesons $(PP)$ and employing dimensional regularization it is given by
\begin{eqnarray}
G_{PP}(s) &= & i \int \frac{d^4 q}{(2\pi)^4} \frac{1}{\left(q^2-m_1^2+i\epsilon\right)\left[(P-q)^2-m_2^2+i \epsilon \right]} \nonumber \\
 & = & \frac{1}{16 \pi^2} \left\{ a(\mu) + \ln \frac{m_1^2}{\mu^2} + \frac{m_2 ^2-m_1^2 + s}{2s} \ln \frac{m_2^2}{m_1^2} \right.\nonumber\\ 
& &  + \frac{p}{\sqrt{s}} \left[
\ln(s-(m_1^2-m_2^2)+2 p\sqrt{s})\right. \nonumber \\ 
& & + \ln(s+(m_1^2-m_2^2)+2 p\sqrt{s}) \nonumber  \\ 
& & - \ln(s-(m_1^2-m_2^2)-2 p\sqrt{s}) \nonumber \\  
& & \left. \left. 
-\ln(s+(m_1^2-m_2^2)-2 p\sqrt{s}) - 2 \pi i \right] 
\right\} \ ,
\label{G2}
\end{eqnarray}
where $ P^2 = s $, $m_1$ and $ m_2$ are pseudoscalar mesons masses, $\mu$ is the regularization energy scale, $a(\mu)$ is a subtraction
constant which absorbs the scale dependence of the integral, and $p$ is the three-momentum in the center of mass frame of the two mesons in channel $PP$,
\begin{eqnarray}
p = \frac{1}{2\sqrt{s}} \sqrt{\left[ s - (m_{1} + m_2 )^2 \right] \left[s - (m_{1} -m_2)^2 \right]} .
\label{pon}
\end{eqnarray}
Notice that if the two-meson loop integral involves a pseudoscalar and a vector meson ($PV$) and two vector mesons ($VV$),  we use standard approximation as in previous studies~\cite{Roca:2005nm,Gamermann:2007fi}, 
\begin{eqnarray}
G_{VP} (s) &= & \left(1 + \frac{p^2}{3 m_1^2 }\right) G_{PP}(s), \nonumber \\
G_{VV} (s) &= & \left(1 + \frac{p^2}{3 m_1^2 }\right) \left(1 + \frac{p^2}{3 m_2^2 }\right) G_{PP}(s).
\label{G3}
\end{eqnarray}
where $m_1$ and $m_2$ represent the masses of vector mesons in the loop. 

%
%
%
%
%
%

The isospin-spin-averaged cross section for the processes in Eq. (\ref{proc1})  are then given in the center of mass (CM) frame  by 
\begin{equation}
\sigma(s) =  \frac{ \chi }{32 \pi s}\overline{\sum_{\text{Isospin}}}\left|\frac{p_f}{p_i}\right||\mathcal{T} (s)|^2 ,
\label{eq:CrossSection}
\end{equation}
\noindent where $p_f$ and $p_i$ are the momentum of the outcoming and incoming particles in the CM frame, respectively; $\chi$ is a constant depending on the total angular momentum of the channel~\cite{Abreu:2018mji}. 
%
%
%
%
%
%
%
%


\section{Results}
\label{Results} 

Now we present numerical results of the cross sections for elastic and inelastic scatterings of $\Upsilon$ by pseudoscalar and vector mesons - i.e. the $\pi,K,\eta,\rho,K^\ast,\omega$ mesons -  using the framework of unitarized coupled-channel $s$-wave amplitudes described in previous section. For convenience, we employ a coupled channel basis by taking into account the quantum numbers $I^G(J^{PC})$, bottom ($\mathbf{B}$) and strangeness ($\mathbf{S}$) of each channel. In Table~\ref{Tab:Canais} is shown the channel content in each sector, obtained by identifying the possible transitions among incoming and outgoing meson pairs with conservation of quantum numbers. 

\begin{table}[h]
	\caption{Channel content in each sector, obtained by identifying the possible $s$-wave transitions among incoming and outgoing meson pairs with conservation of quantum numbers. }
			\begin{tabular}{c|c}
		\hline
		$\mathbf{I^G(J^{PC})}$& $\mathbf{B=0, S=0}$ \\ 
		\hline
				\hline

		$0^+(0^{++}), 0^-(1^{+-})$ 		& \multirow{2}{*}{$\Upsilon \Upsilon , \Upsilon  \omega, \omega \omega, \rho \rho, B_s^\ast \bar B_s^\ast$}\\
		$0^+(2^{++})$ 					& \\ 
		\hline
		\multirow{2}{*}{$0^-(1^{+-})$}	& $\pi \rho, \eta \omega, \Upsilon \eta , \eta_b \omega, K \bar K^\ast - c.c.,$\\
		&  $ \Upsilon \eta_b, B \bar B^\ast - c.c., B_s \bar B_s^\ast + c.c.$ \\ 
		\hline
		\multirow{2}{*}{$1^-(0^{++})$}	& $\rho \omega, K^\ast \bar K^\ast, \eta \pi, \bar K K$ \\
		& $ \Upsilon \rho , B^\ast \bar B^\ast, \eta_b \pi, \bar B B$ \\
		\hline
		$1^+(1^{+-}), 1^-(2^{++})$		& $\Upsilon \rho , \rho \omega, K^\ast \bar K^\ast, B^\ast \bar B^\ast$\\ 
		\hline
		\multirow{2}{*}{$1^+(1^{+-})$}	& $\pi \omega, \eta \rho, K \bar K^\ast + c.c.$\\
		& $\Upsilon \pi , \eta_b \rho, B \bar B^\ast + c.c.$\\
		\hline 	
		\hline
		$\mathbf{I^G(J^{PC})}$	& $\mathbf{B=0, S=1}$\\ 
		\hline
		\hline
      	\multirow{2}{*}{$\frac{1}{2}(0^+)$}		& $K \eta, K \pi, K^* \omega, K^* \rho$ \\
		& $K \eta_b, B_s \bar B, \Upsilon  K^* , B_s^\ast \bar B^\ast$ \\
		\hline
		\multirow{1}{*}{$\frac{1}{2}(1^+),\frac{1}{2}(2^+)$} 			& $\Upsilon K^* , K^* \omega, K^* \rho, B_s^\ast \bar B^\ast$ \\ 
%
		\hline
		\multirow{2}{*}{$\frac{1}{2}(1^+)$}		& $\pi K^\ast, \eta K^\ast, K \rho, K \omega$\\
		& $\eta_b K^\ast,\Upsilon  K,\bar B B_s^\ast, \bar B^\ast B_s$\\	\hline
	\end{tabular}	
	\label{Tab:Canais}
\end{table}

We have employed in the calculations the isospin-averaged meson masses:  $m_{\pi} = 138 $ MeV, $m_{\rho} = 771 $ MeV, $m_K = 495 $ MeV, $m_{\eta} = 548$ MeV, $m_{\omega} = 782 $ MeV, 
$m_{K^{\ast}} = 892 $ MeV, $m_B = 5279 $ MeV, 
$m_{B^{\ast}} = 5324$ MeV, $m_{B_s} = 5308 $ MeV, $m_{B_s ^{\ast}} = 5355$ 
MeV, $m_{\eta_b} = 9399$ MeV,  $m_{\Upsilon} = 9460 $ MeV, $m_{L} = 800 $ MeV, $m_{H} = 5000 $ MeV and $m_{H}^\prime = 9000 $ MeV; for the decay constants: $f_{\pi} = 93$ MeV and  $f_{B} = 189.9$ MeV~\cite{Bazavov:2017lyh}. 

As a free parameter, the subtraction constant $a (\mu)$ has been fixed following Ref.~\cite{Dias:2014pva} (see also \cite{Abreu2013a} for a detailed explication): by matching the loop function obtained via dimensional regularization, Eq.~(\ref{G2}), to the one in cutoff regularization scheme. 
Then, its fitted value corresponding to a reasonable cutoff momentum of about 700~MeV  GeV is $a = - 3.78 $ at the scale $ \mu = 1.5 $ GeV, which has obtained from the matching of the elastic $\Upsilon \pi$ channel. In principle, a fine tuning in the subtraction constants might be performed for each channel to replicate the experimental data (if they were available). But we have noted from the test with other possible values of $a (\mu)$ the occurrence of some artificial structures (e.g. dip or cusp structures) in the region of CM energy below the $B_{(s)}^{(\ast)} \bar B_{(s)}^{(\ast)}$ threshold of a specific channel, and no qualitative changes above it.  
In this sense, bearing in mind the aim of the present study, that is to provide an overview on the quarkonia interactions with surrounding medium, as it will be seen the chosen free parameter gives rise to cross sections without such mentioned structures below the cited thresholds. Indeed, a cusp effect will appear only at the respective threshold, induced by the opening of the $B_{(s)}^{(\ast)} \bar B_{(s)}^{(\ast)}$ channel, in general not corresponding to a resonant state. Hence, we believe that this approach will provide at least a reasonable understanding of the behavior of the studied reactions.

\begin{figure}
	\centering
	\includegraphics[width=1.0\linewidth]{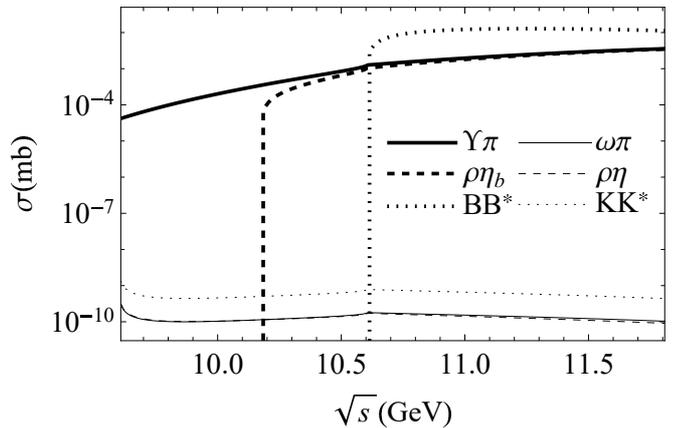}
	\caption{Cross sections for $\Upsilon \pi$ scattering into allowed final states as a function of the CM energy $\sqrt{s}$. }
	\label{fig:RelevanceYpi}
\end{figure}

Because of its large multiplicity, the processes involving pions in initial state should yield the most relevant contributions in relation to other light mesons. For that reason, we begin with the analysis of the cross sections for $\Upsilon \pi$ scattering into the final states consistent with the conservation of quantum numbers: $ (B \bar B^\ast + c.c.), \Upsilon \pi, \rho \eta_b, \omega \pi, \rho \eta, ( \bar K^\ast K+c.c. )$. They are displayed in Fig.~\ref{fig:RelevanceYpi}. 
Some of them are not analyzed in previous studies using different frameworks, e.g. Refs.~\cite{Lin:2000ke,Abreu:2018mnc}. 
One important feature of this model is that the only non-vanishing cross section at tree level is the one related to the process with final state $( B \bar B^\ast + c.c.)$, presenting an almost linear growth at higher CM energies. 
When the unitarized amplitudes are employed, the cross sections experience significant changes. In the case of the channel $( B \bar B\ast + c.c.)$, it becomes smaller by about one order of magnitude at  $\sqrt{s} = 11$ GeV with respect to non-unitarized situation, with a peak near the threshold and decreasing as the energy grows. Moreover, the other processes acquire nonzero cross sections through  meson loops. 
However, as expected the significant processes are those whose final states carrying mesons with bottom quarks, i.e. $ (B \bar B^\ast + c.c.), \Upsilon \pi, \rho \eta_b$, with the $B \bar B^\ast$ channel reaching the cross section with higher magnitude. On the other hand, not surprisingly the $\omega \pi, \rho \eta, ( \bar K^\ast K+c.c. )$ channels give contributions smaller than the bottomed channels by about a factor $10^{-6}-10^{-7}$, which legitimates to neglect them for pragmatic goals.

\begin{figure}
	\centering
	\includegraphics[width=1.0\linewidth]{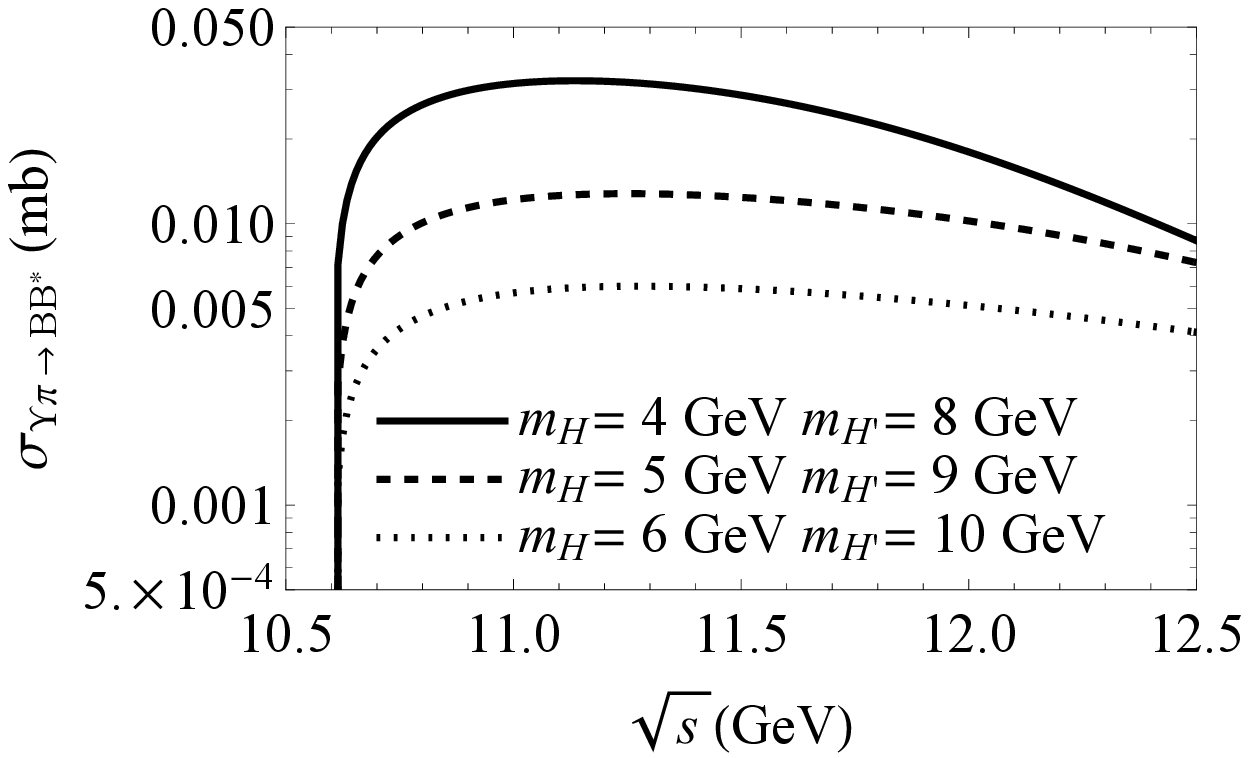} \\	\includegraphics[width=1.0\linewidth]{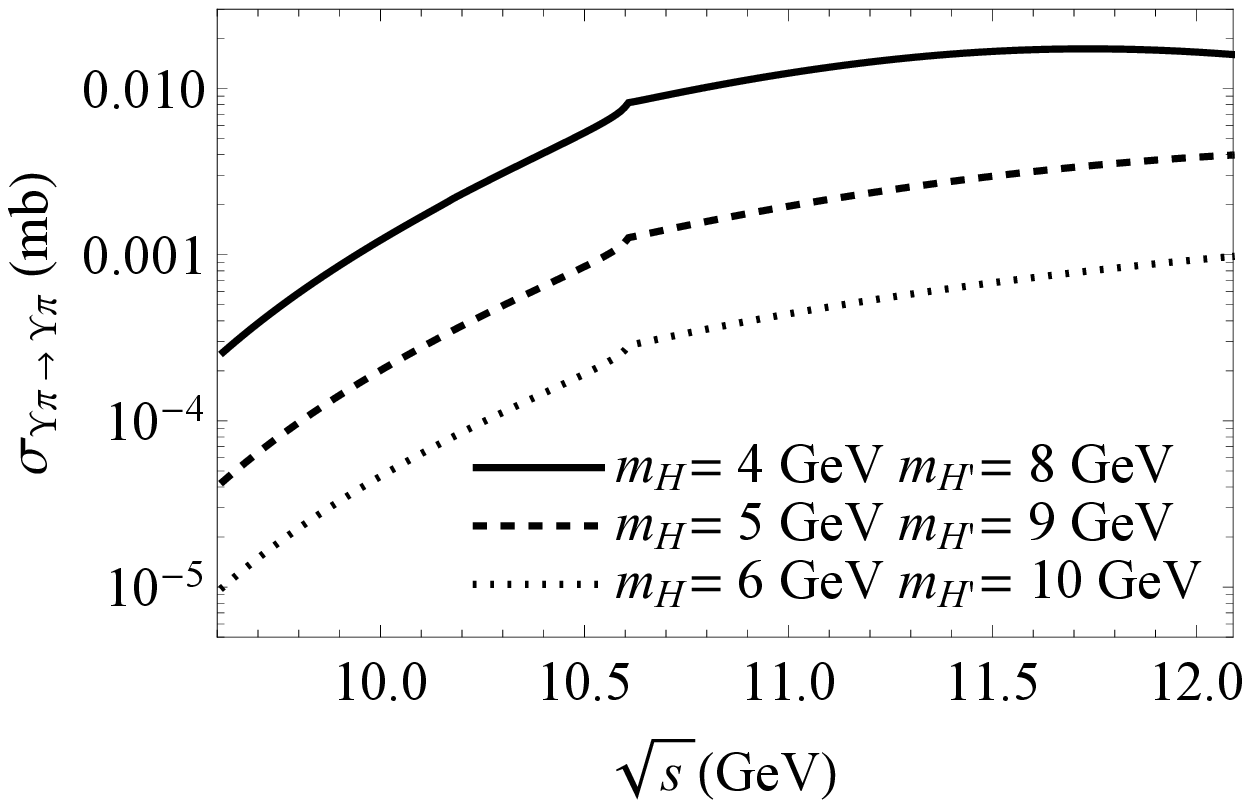}\\ 
	\includegraphics[width=1.0\linewidth]{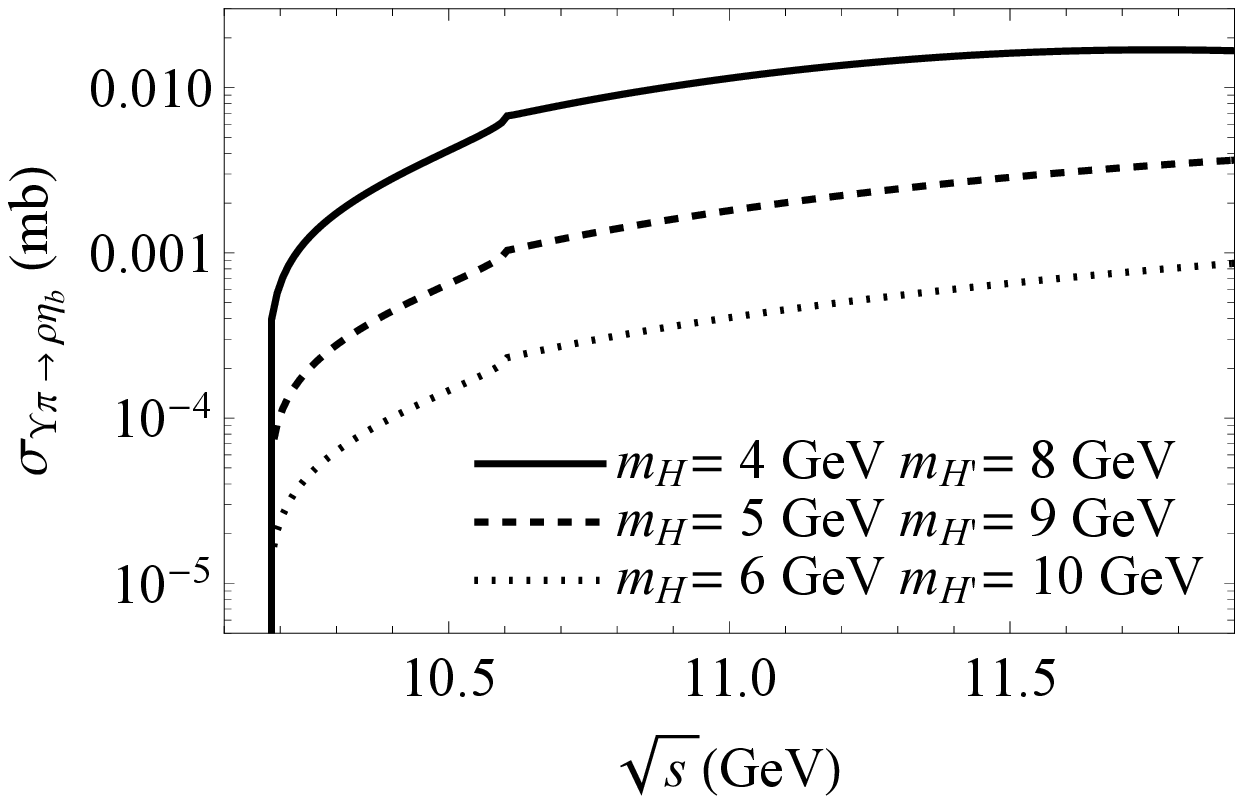} 
	\caption{Cross sections for $\Upsilon \pi \rightarrow (B \bar B^\ast + c.c.), \Upsilon \pi, \rho \eta_b$ reactions as a function of the CM energy $\sqrt{s}$, taking different values of the parameters $m_L, m_H$ and $m_H^{\prime}$. }
	\label{fig:RelevanceYpi2}
\end{figure}

Another analysis performed is the dependence of the relevant channels taking different values of the parameters $m_L, m_H$ and $m_H^{\prime}$. Thus, from the plots in Fig.~\ref{fig:RelevanceYpi2} we estimate the modifications on the magnitudes of the cross sections in the considered ranges of the parameters. The augmentation of $ m_H$ and $m_H^{\prime}$ (mainly $ m_H$) induces smaller cross sections.

Our results for the channel $\Upsilon \pi \rightarrow (B \bar B^\ast + c.c.)$ can be contrasted to the ones of preceding works~\cite{Lin:2000ke,Abreu:2018mnc}, obtained employing different approaches. 
Taking the respective maximum values, this cross section in Fig.~\ref{fig:RelevanceYpi} is appreciably smaller than the ones reported in the mentioned Refs. As follows we will try to shed light on the causes of this discrepancy. 
Firstly, the dissimilar energy dependence between the formalisms plays a substantial role on this issue. In the present framework the $s$-dependence is encoded in the unitarized amplitude, extracted from the Bethe-Salpeter equation, through the $s$-wave projected potential $V(s)$ and the loop function $G(s)$. A limiting aspect is that higher partial waves contributions have not been included. Apart from that, the high-energy behavior is controlled on more rigorous grounds, without invoking any ad hoc hypotheses. Having said that, it should be observed that the change of the cutoff and subtraction constant may cause noticeable modifications in the magnitudes and shapes of the cross sections~\cite{Dias:2014pva}, which is useful to reproduce the experimental data when accessible. Diversely, Refs.~\cite{Lin:2000ke,Abreu:2018mnc} make use of distinct effective Lagrangians and the form-factor approach, which is responsible for the high energy behavior of scattering amplitudes. Its choice is arbitrary, except in some specific cases, in which the form factor can be derived (for example via QCD sum rules). As a consequence, the type of the form factor (as for instance monopole-, dipole- or Gaussian-type ones) and the value of cutoff can alter the shape and magnitude of the cross sections by orders of magnitude,  as recognized by~\cite{Lin:2000ke,Abreu:2018mnc}, which might bring their findings closer to the ones reported here. 

Secondly, the coupling strengths provide obviously a notable contribution on this matter. 
Here they are given by the constants $f_{\pi}$, $f_{B}$ and the suppression factors $\gamma$ and $\psi$. 
 As a test, we have checked that the utilization of $\psi = (m_L/ m_H ^{\prime})^2$, as in~\cite{Dias:2014pva}, augments the cross sections by a factor of about one order of magnitude. Additionally, regarding the decay constant $f_{B}$, it can be seen from diverse lattice-QCD computations  that its isospin-averaged value can vary between 160 and 250 MeV, including the current errors~\cite{Aoki:2016frl}. Then, we have verified an alteration of the cross section by a factor of $20\%$ using these limits for $f_{B}$.
Contrastingly, the cited Refs. have made use of other assumptions: Ref.~\cite{Lin:2000ke} relates the coupling constants to the $SU(5)$ universal coupling constant and to those involving light and charm mesons via the vector-meson dominance model. In the case of~\cite{Abreu:2018mnc}, a $SU(4)$ effective formalism in which the vector mesons are identified as the gauge bosons has been employed, including anomalous parity interactions, with the coupling constants determined from the generalization of the relations of the charm sector also based on the vector-meson dominance model. The reader is also invited to consult the paper~\cite{Abreu:2017nuc} for another illustration on the determination of the coupling constants, founded on the experimental information from the charm sector and from heavy quark symmetry. 


\begin{figure}
	\centering
	\includegraphics[width=1.0\linewidth]{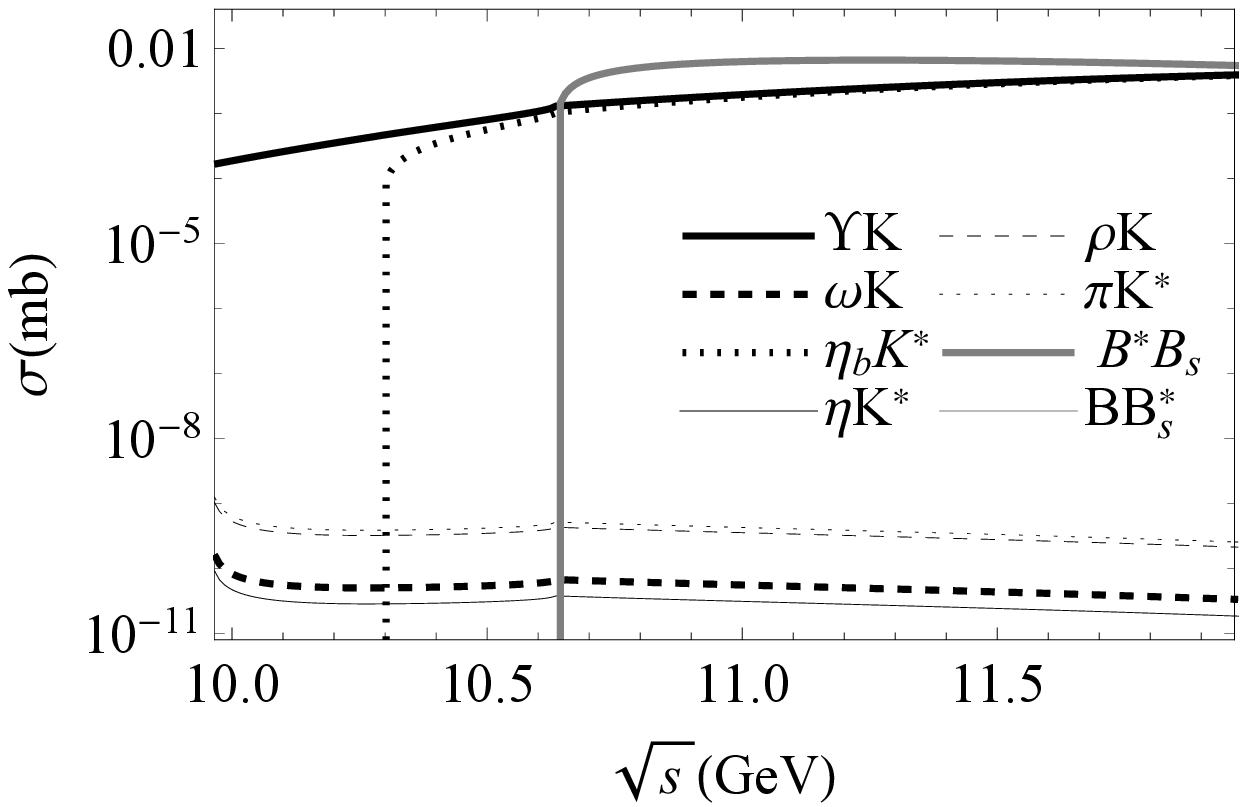} \\ 
	\includegraphics[width=1.0\linewidth]{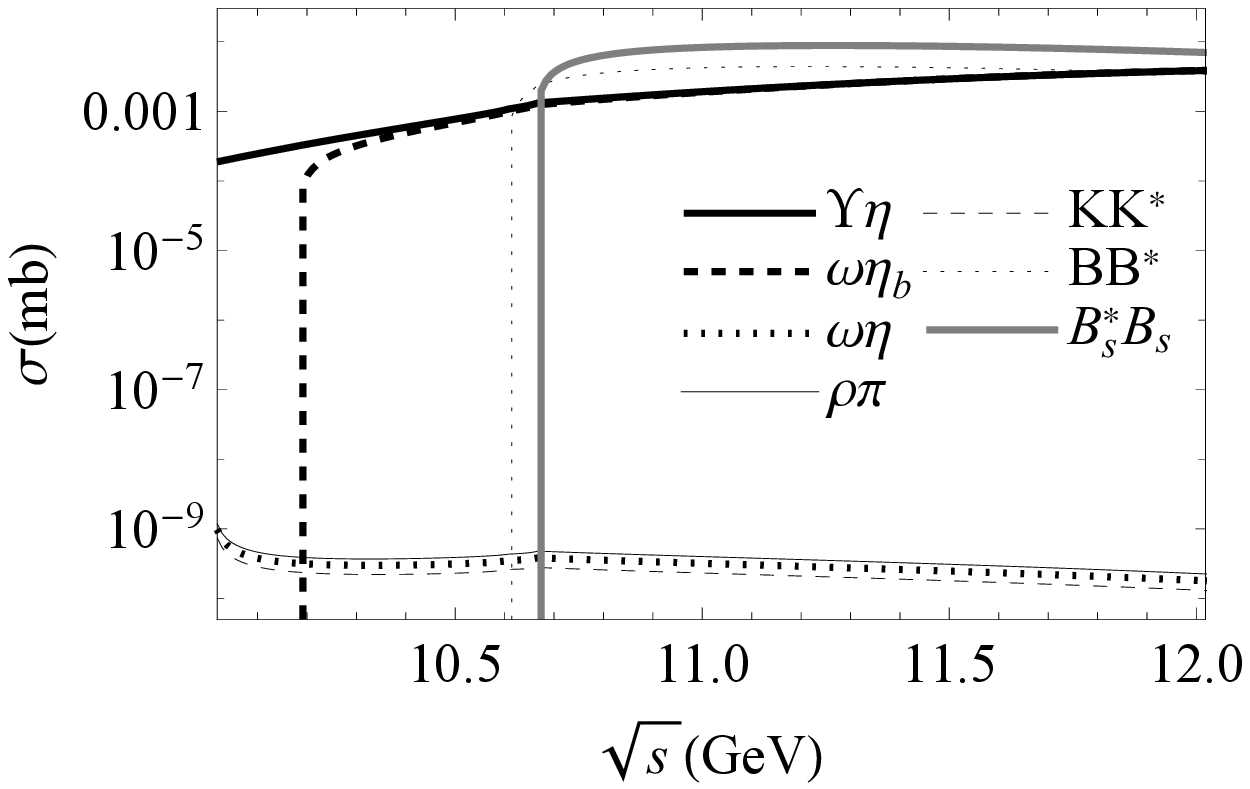}
	\caption{Unitarized cross sections for $\Upsilon K $ (top panel) and $\Upsilon \eta $ (bottom panel) scatterings into allowed final states as a function of the CM energy $\sqrt{s}$. }
	\label{fig:RelevanceYKeta}
\end{figure}

\begin{figure}
	\centering
	\includegraphics[width=1.0\linewidth]{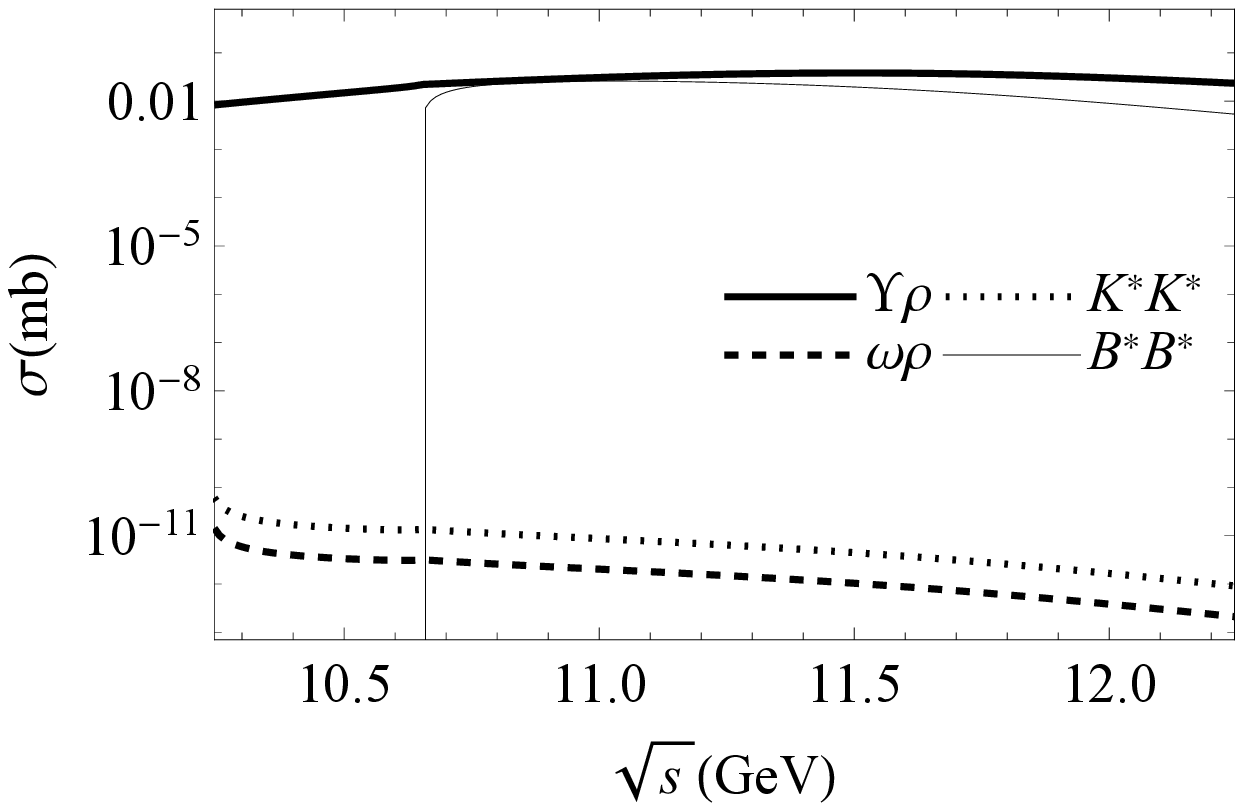} \\ 
	\includegraphics[width=1.0\linewidth]{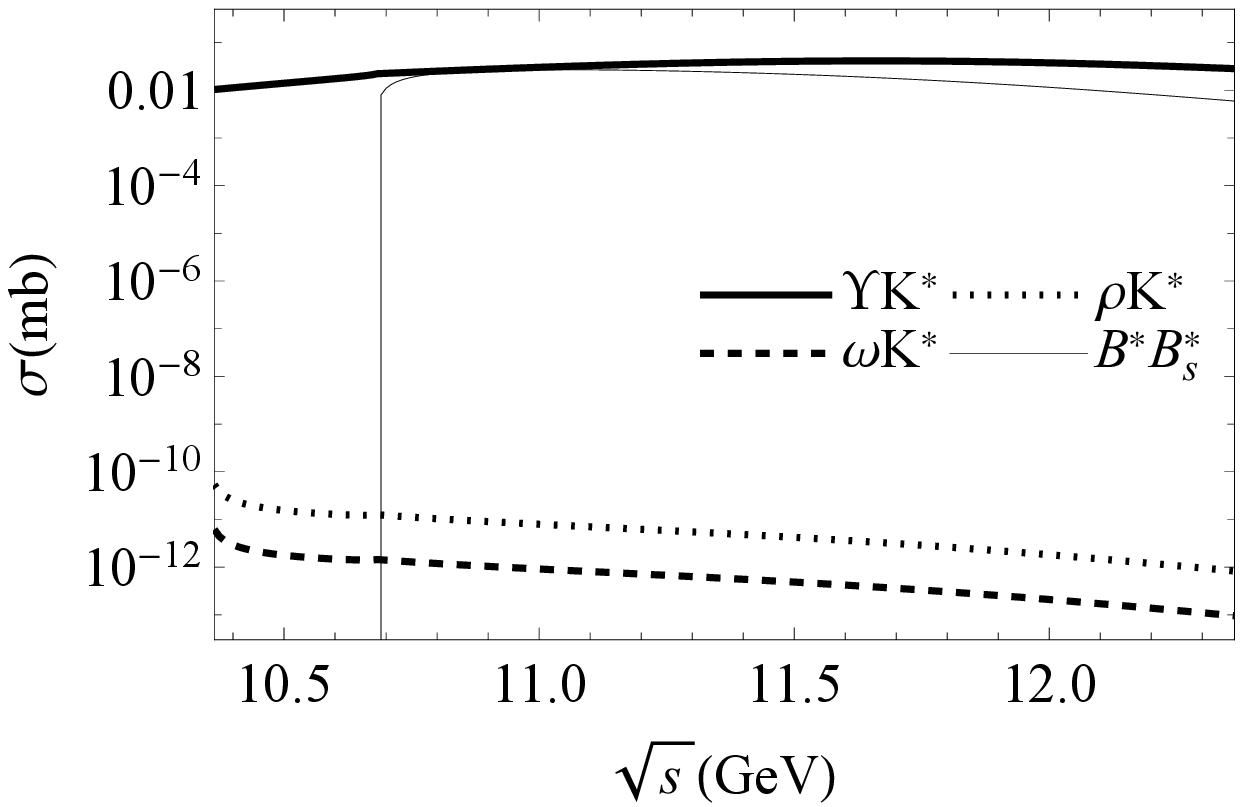} \\
		\includegraphics[width=1.0\linewidth]{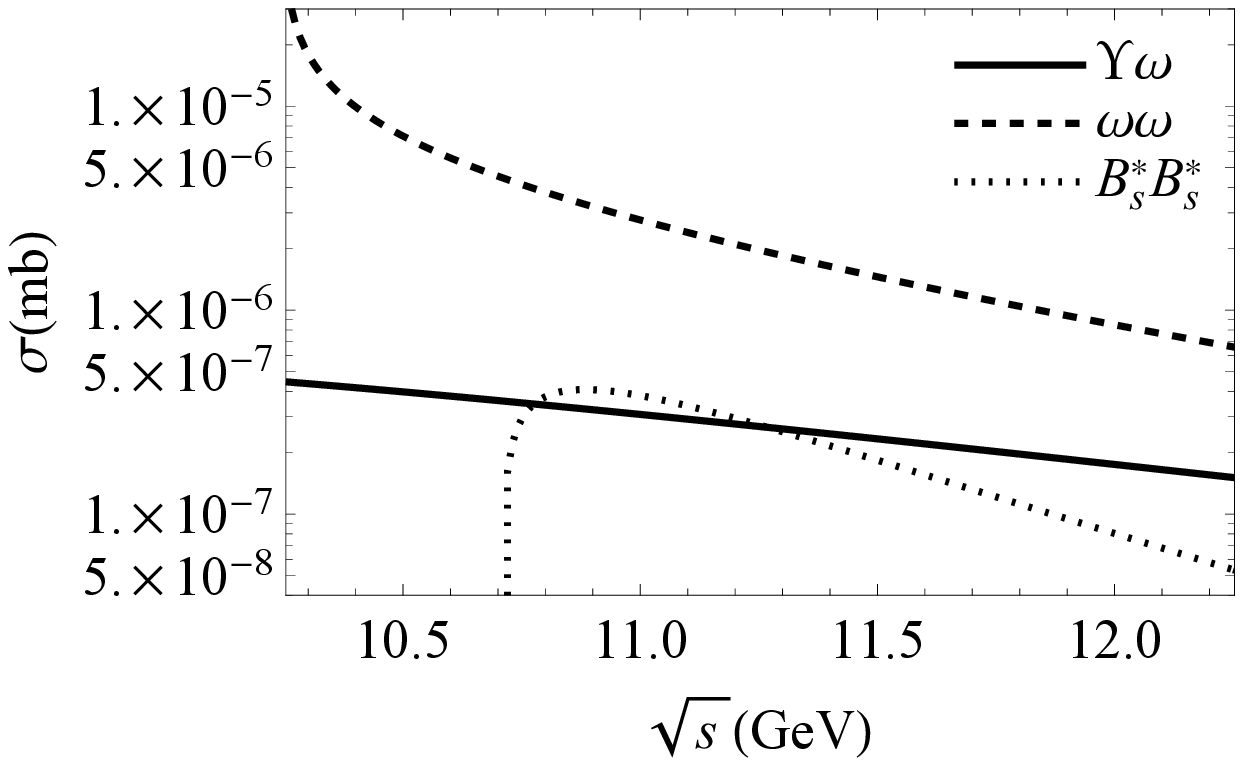}
	\caption{Unitarized cross sections for $\Upsilon \rho $ (top panel), $\Upsilon K^{\ast} $ (center panel), and $\Upsilon \omega $ (bottom panel) scatterings into allowed final states as a function of the CM energy $\sqrt{s}$. }
	\label{fig:RelevanceYrhoKstaromega}
\end{figure}

The unitarized cross sections for $\Upsilon X$ scatterings into the most relevant final states, with $X$ being the $K, \eta,\rho, K^{\ast} , \omega $ mesons, are also plotted in Figs.~\ref{fig:RelevanceYKeta} and ~\ref{fig:RelevanceYrhoKstaromega}. We remark the following general properties: at tree level only reactions with open bottomed mesons in final states engender non-zero cross sections, with an uncontrolled behavior at higher energies. But the unitarized coupled channel amplitudes get non-vanishing cross sections via the meson loops, with a controlled decrease as energy grows. For all cases of $X = K, \eta,\rho, K^{\ast} , \omega $, the processes with bottomed mesons or a bottomonium in the final state are the most relevant contributions for the cross sections, whereas the other contributions have negligible magnitudes and are strongly suppressed. 
Focusing on the absorption of $\Upsilon $ by vector mesons, specifically the $\rho$ meson, we see that the channels with bottomed mesons have a cross section comparable or even greater than the $\sigma_{\Upsilon \pi \rightarrow (B \bar B^\ast + c.c.)}$ in Fig.~\ref{fig:RelevanceYpi}. This is in contrast with the findings in \cite{Lin:2000ke}, calculated in a different framework as emphasized above. The possible explanation on this point comes from the amplitudes in Eqs.~(\ref{Eq:CasoVPVP})-(\ref{Eq:CasoVVVV}) engendered by the interactions: while the processes $VP \rightarrow VP $ occur only via $s$-channels, the $VV \rightarrow VV $ reactions happen via $s,t,u$-channels. Besides, here the interactions are described by contact vertices, with the coupling strengths depending on the same constants ($f_{\pi}$, $f_{B}$, $\psi$ and $\gamma$), whereas in \cite{Lin:2000ke} there are distinct contributions from contact and one-boson exchange diagrams for the amplitudes.

Other point worthy of mention is that in the plots of the cross sections for $\Upsilon $ scattering by vector mesons, we have considered the sum of the reactions with different spin contributions $ J = 0,1,2 $), but considering only the $ VV  \rightarrow VV $ processes, because of the large suppression of $ VV \rightarrow PP $ ones. As discussed in Ref.~\cite{Abreu:2018mnc}, since $VV \rightarrow PP$ occurs via $s$-channel, and $(t-u)$ present in Eq.~(\ref{Eq:CasoVVPP}) gives $(m_1^2-m_2^2)(m_1^{'2}-m_2^{'2})$, where $m_i$ are the masses of the incoming particles and $m_i'$ the masses of the outgoing particles, when $m_1 \sim m_2$ or $m_1^{'} \sim m_2^{'}$ it becomes negligible.

  \begin{center}
	\begin{figure}[h]
		\centering
		\includegraphics[width=1.0\linewidth]{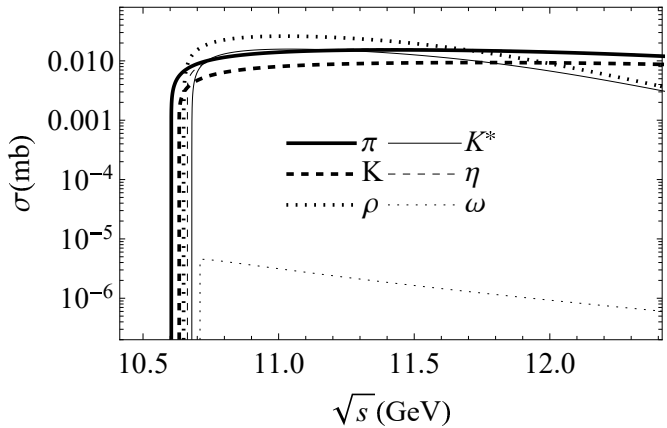}\\
		\caption{Cross-sections as function of center-of-mass energy $\sqrt{s}$ for $\Upsilon X $ scattering into all allowed final states; $X$ denotes $\pi,K,\eta,\rho,K^\ast,\omega$ mesons. }
		\label{fig:Relevance_IN}
	\end{figure}
\end{center}
%
%
%
%
%
%
%

We complete this section with the plots in Fig.~\ref{fig:Relevance_IN} of the cross-sections for the $\Upsilon X$ into all possible channels according to Table~\ref{Tab:Canais}). Near the thresholds, the cross-sections involving the $\rho, K^*$ mesons have comparable or even greater magnitudes than those with other mesons. 


\section{Finite-temperature effects}
\label{Finite-temperature effects} 

The evaluation performed above has taken into account the bottomonium dissociation reactions in vacuum, which is in principle useful in the estimation of the $\Upsilon $ suppression in nucleus-nucleus collisions. However, several studies pointed out that hadrons have their masses affected by the medium, in particular those relevant in heavy-ion-collision phenomena~\cite{Gao:2019idb,Zhou:2012vv,Ji:2015fva,Cleven:2017fun,Gu:2018swy,Montana:2020lfi}. In this sense, we perform a brief analysis of finite-temperature effects on the $\Upsilon X$-absorption cross sections. At our best knowledge, this has not been studied either. 

To this end, we exploit Imaginary Time Formalism (ITF), in which the propagators suffer finite-temperature modifications whereas the couplings stay unaltered. However, instead of a more rigorous and complex framework, which would require the dressing of the meson propagators
with the self energy (see e.g. Ref.~\cite{PhysRevC.96.045201}), as a first attempt we exploit a simple and naive way to investigate the impact of finite temperature: we replace the loop function $G_{PP}(s)$ in Eq.~(\ref{G2}) by its finite-temperature version, $G_{PP}^{(FT)}(s)$, which can be written as~\cite{LeBellac,Gao:2019idb}, 
\begin{eqnarray}
G_{PP}^{(FT)}(s) = G_{PP}(s) + G_{PP}^{(T \neq 0)}(s),  
\label{GFT1}
\end{eqnarray}
where $G_{PP}^{(T \neq 0)}(s)$ denotes the finite-temperature corrections and is given by
\begin{widetext} 
\begin{eqnarray}
G_{PP}^{(T \neq 0)}(s) &= &  \int^{\infty}_{0} \frac{k^2 dk}{8\pi^2  E_1 E_2} \left\lbrace \frac{1}{\sqrt{s} + E_1 + E_2} \left[ f(E_1) + f(E_2)\right]
 + \frac{1}{\sqrt{s} + E_1 - E_2} \left[-f (E_1) + f(E_2)\right]\right. \nonumber \\
\nonumber & & + \left. \frac{1}{\sqrt{s} - E_1 + E_2} \left[f(E_1) - f(E_2) \right] \right\rbrace - P.V. \int^{\infty}_{0} \frac{k^2 dk}{8\pi^2 E_1 E_2} \frac{1}{\sqrt{s} - E_1 - E_2} \left[f (E_1) + f(E_2) \right]\\
& & + \frac{iq(s)}{8\pi \sqrt{s}} \left[f(\tilde{E}_1) + f(\tilde{E}_2)\right] \theta (s-s_{th}), \label{GFT2}
\end{eqnarray}
\end{widetext}
with $\beta = 1/T$; $E_i = \sqrt{k^2 + m_i^2}$; $f(x)$ being the Bose distribution function, i.e. $f(x) = (\exp{(\beta x)} - 1)^{-1}$; and $\tilde{E}_i = \sqrt{p(s)^2 + m_i^2}$, $p(s)$ representing the magnitude of the on-shell three momenta in the CM frame, defined in Eq.~(\ref{pon}). 
As a consequence, the matrix representing the unitarized coupled channel transitions at finite temperature $T$ reads
\begin{eqnarray}
  \mathcal{T}^{(FT)}(s) = \frac{V(s)}{1+V(s)G_{PP}^{(FT)}(s)}. 
  \label{T}
\end{eqnarray}

\begin{center}
\begin{table}[h]
\caption{
Input thermal masses for increasing temperatures (All in GeV). The behaviour of thermall masses for the $\pi,K,\eta$ mesons are mattter of debate, so our choice is the use of results from Ref.~\cite{Gu:2018swy}, which give masses almost constant. Concerning other relevant mesons, we are not aware of detailed studies on them; we adopt the parametrization presented in~\cite{Zhou:2012vv}, 
with critical temperatue $T_c = 0.175 $ GeV 
. }
\vskip1.5mm
\label{thermal-masses}
\begin{tabular}{c|c|c|c|c|c}
\hline
\hline
T & $ 0$ & $ 0.05$ & $ 0.1$ & $ 0.15$ & $ 0.16$   \\ 
\hline
$\pi$ & 0.13499 & 0.13515 & 0.13633 & 0.13894 & 0.13962\\
$K$ & 0.49421 & 0.49427 & 0.49437 & 0.49531 & 0.49569\\
$\eta$ & 0.55138 & 0.55134 & 0.55113 & 0.55299 & 0.55383\\  
$\rho$ & 0.77534 & 0.76738 & 0.67397 & 0.32494 & 0.20361\\
$K^*$ & 0.89302 & 0.88904 & 0.82195 & 0.50912 & 39089\\
$\omega$ & 0.78265 & 0.77815 & 0.74145 & 0.59990 & 0.53324\\
$\eta_b$ & 9.39870 & 9.39624 & 9.36353 & 9.18255 & 9.07648\\
$D$ & 1.8650 & 1.8650 & 1.86318 & 1.7728 & 1.69257\\
$D*$ & 2.0080 & 2.00722 & 1.97546 & 1.7297 & 1.62074\\
$D_s$ & 1.9680 & 1.96562 & 1.94299 & 1.84189 & 1.78874\\
$D^*_s$ & 2.1120 & 2.11175 & 2.09562 & 1.90527 & 1.80378\\
$B$ & 5.27925 & 5.27924 & 5.27409 & 5.01826 & 4.79129\\
$B*$ & 5.32470 & 5.32262 & 5.23841 & 4.58671 & 4.29789\\
$B_s$ & 5.36684 & 5.36035 & 5.29862 & 5.02271 & 4.82506\\
$B^*_s$ & 5.41580 & 5.41526 & 5.37382 & 4.88568 & 4.57359\\
$J/\psi$ & 3.09700 & 3.09619 & 3.08541 & 3.02578 & 2.99073\\
$\Upsilon$ & 9.46030 & 9.45783 & 9.42490 & 9.24273 & 9.13567\\
\hline
\hline
\end{tabular}
\end{table}
\end{center}

The other ingredient in the present approach is the inclusion of the thermal corrections to the masses of the mesons according to Table~\ref{thermal-masses}.  We are aware of the possible controversies on this choice of parametrization (see for example discussions in Refs.~\cite{Cleven:2017fun,Gu:2018swy,Montana:2020lfi,Zhou:2012vv,Ji:2015fva,Gu:2018swy}, and the particular case of pion in~\cite{Schenk:1993ru,Nicola:2014eda,Cheng:2010fe}). But our point here is to estimate the modifications of the threshold effects contained in the $s$-channel unitarity loop functions, caused by the shifts of the thermal masses. This in principle might engender changes in the determination of the thermal
properties of the cross sections.



But before considering the finite-temperature effects on the cross sections for the $\Upsilon X$ dissociation reactions, for completeness we also analyze the reactions involving the charmonium absorption by light mesons, discussed in Ref.~\cite{Abreu:2018mnc} without thermal effects. 

\begin{widetext} 

\begin{figure}
	\centering
	\includegraphics[width=5.9cm]{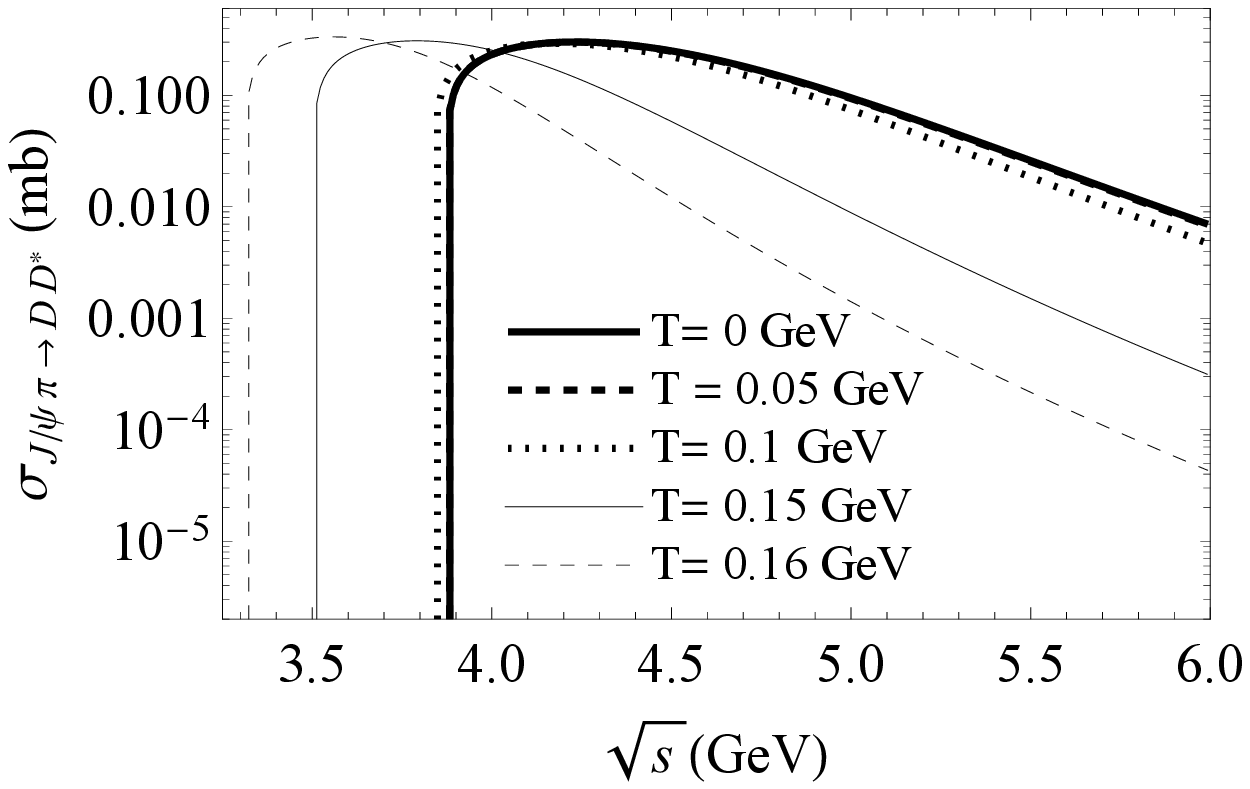}
	\includegraphics[width=5.9cm]{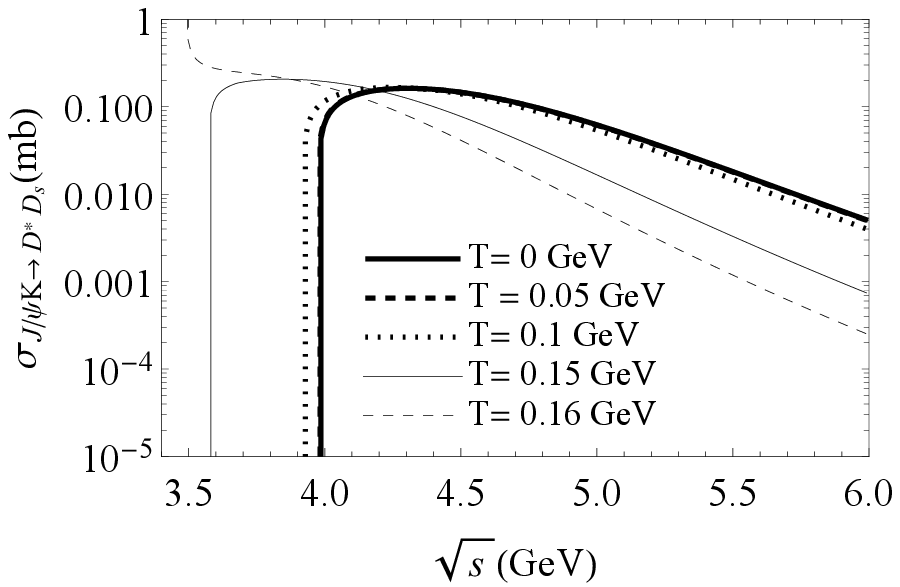} 
	\includegraphics[width=5.9cm]{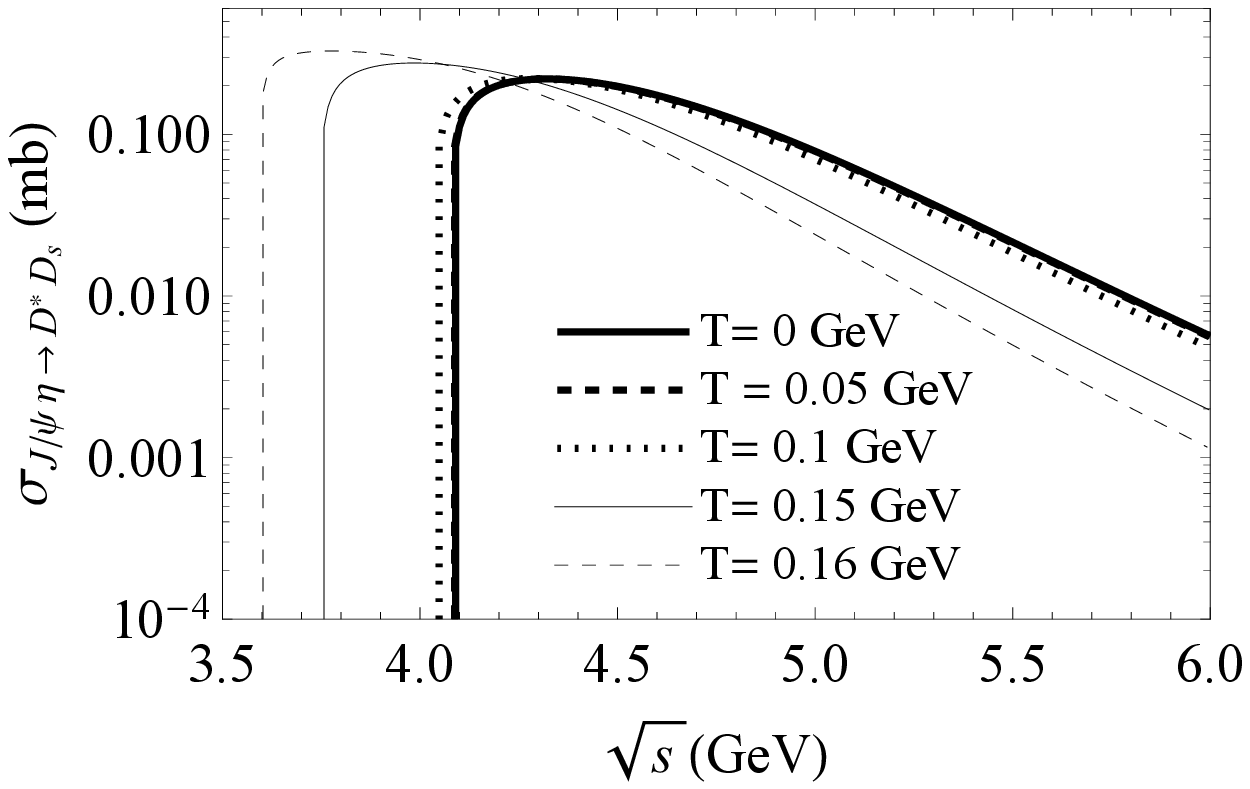}
	\\
	\includegraphics[width=5.9cm]{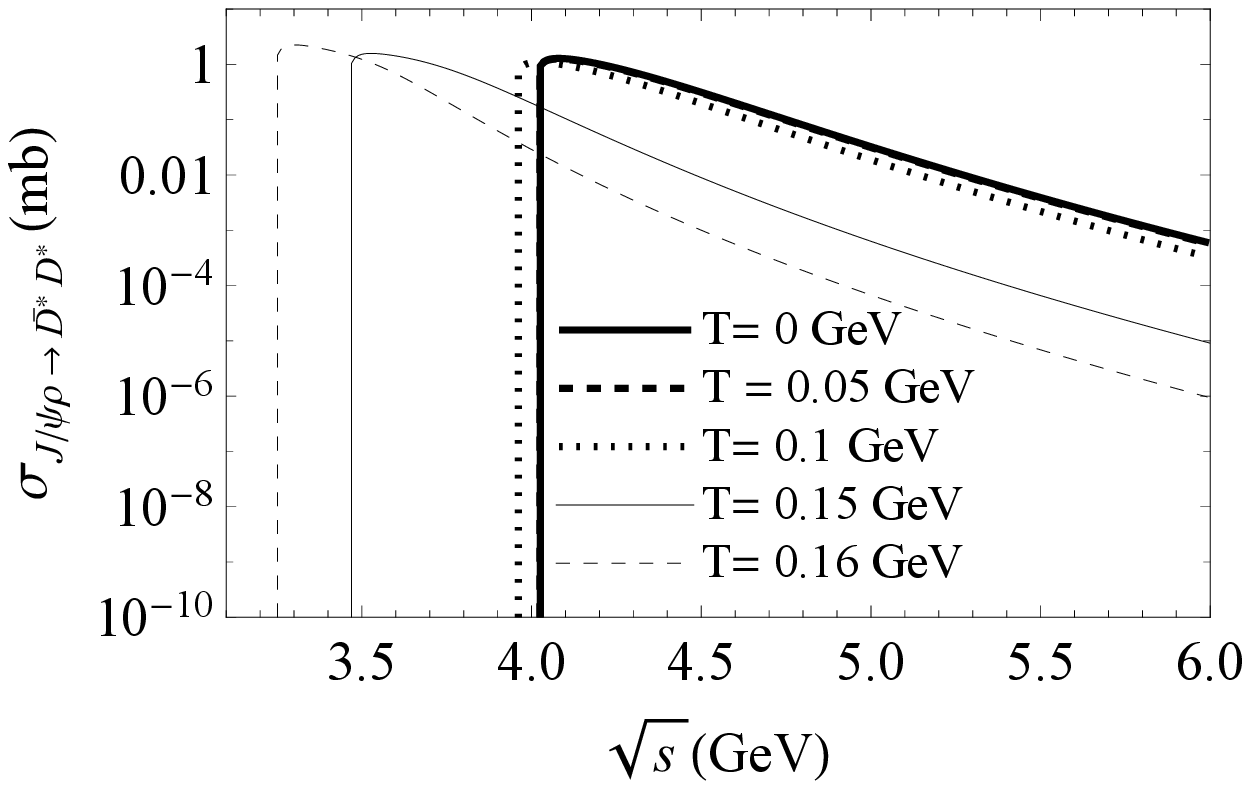}  
	\includegraphics[width=5.9cm]{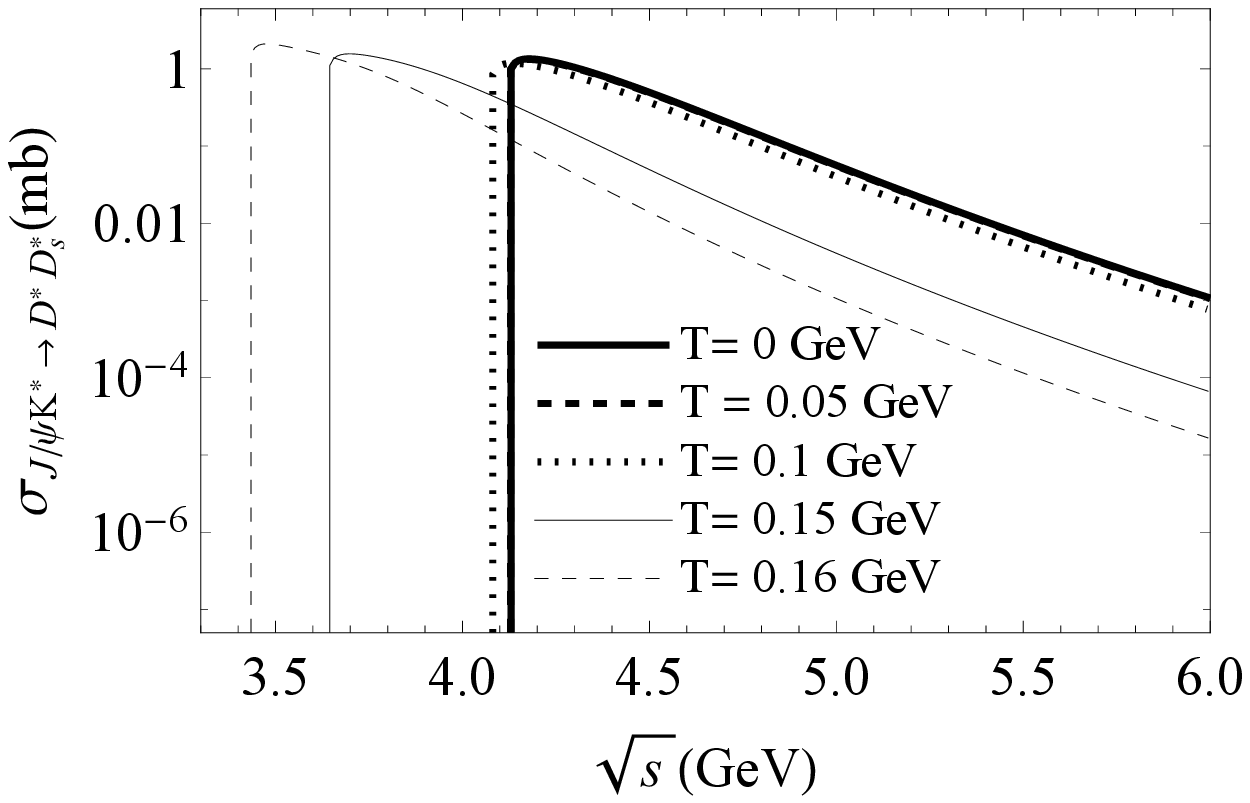} 
	\includegraphics[width=5.9cm]{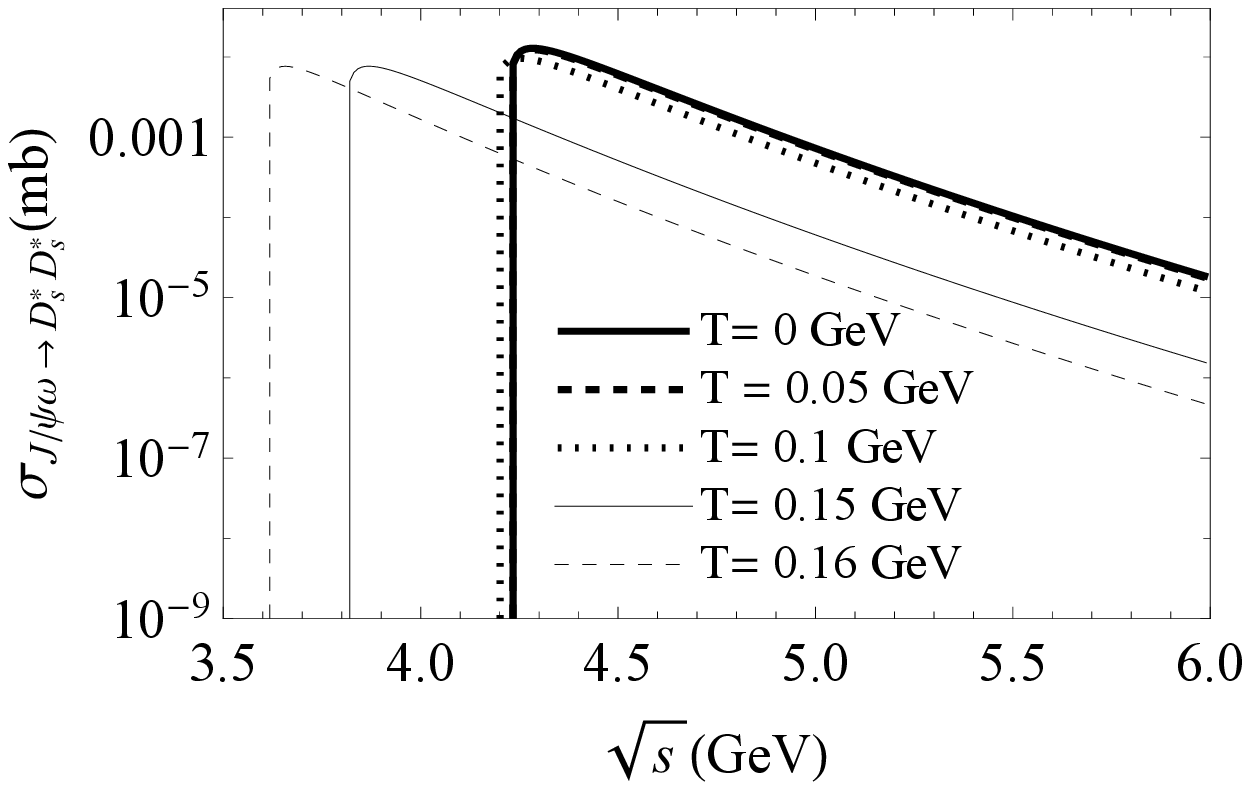}
	\caption{Unitarized cross sections for $J/\psi X  (X=\pi,K,\eta,\rho,K^\ast,\omega)$   scatterings into relevant final states with open-charm mesons as a function of the CM energy $\sqrt{s}$, at different values of temperature. The curves corresponding to $T=0.05$ MeV almost coincide with those with $T=0$ MeV. }
	\label{fig:RelevanceJPsiXFT}
\end{figure}

\end{widetext}

The $J/\Psi X$ reactions are displayed in Fig.~\ref{fig:RelevanceJPsiXFT}, taking into account only the final states with open-charm mesons, which are the most relevant and representative channels. As the temperature increases, the peak of the cross sections is affected, 
moving to the left due to the reduction of the threshold of the respective channel.  
For a certain range of temperature, this general behavior is qualitatively similar to the one reported in Ref.~\cite{Zhou:2012vv}, in which the cross sections are obtained by employing the quark-interchange mechanism, the Born approximation, and a temperature-dependent quark potential. But we remark that a direct quantitative comparison is tricky, due to the different values and behaviors of thermal masses and range of temperature, which in ~\cite{Zhou:2012vv} it is given in terms of the critical temperature.

\begin{widetext} 

\begin{figure}
	\centering
	\includegraphics[width=5.9cm]{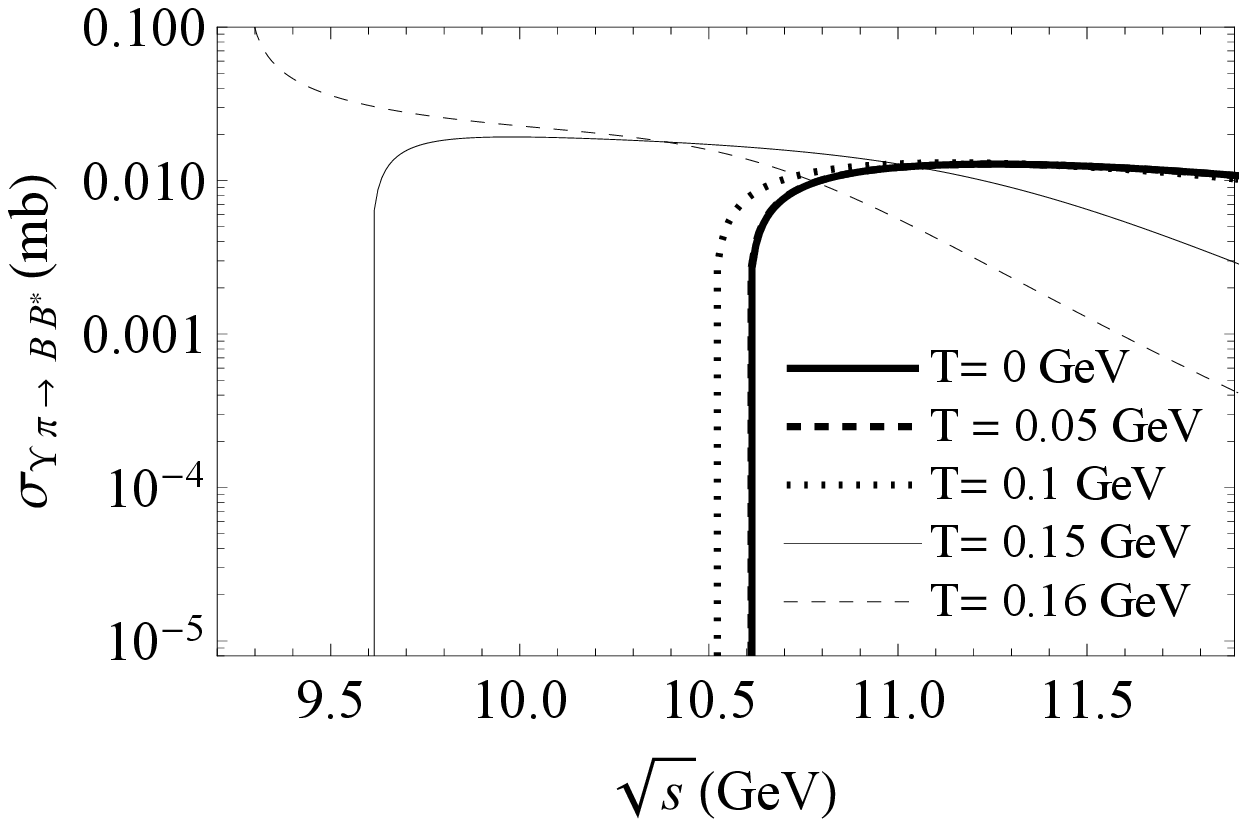}
	\includegraphics[width=5.9cm]{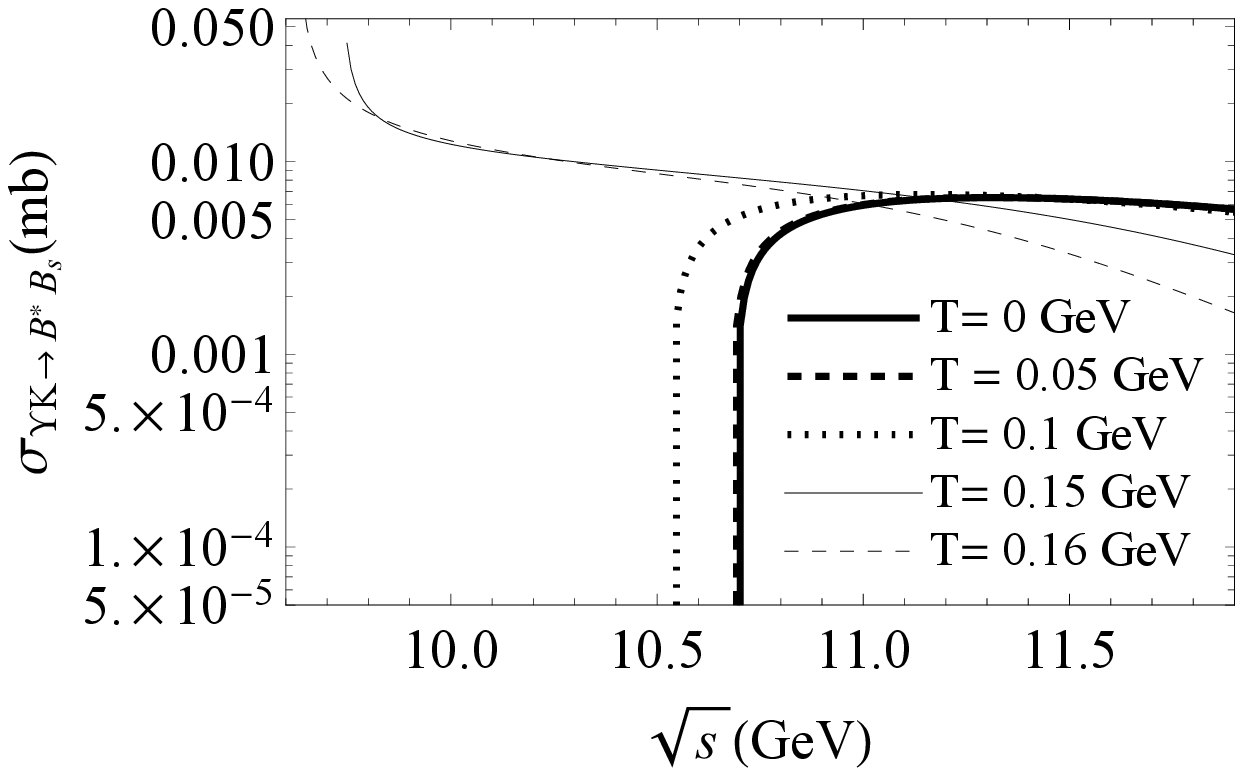} 
	\includegraphics[width=5.9cm]{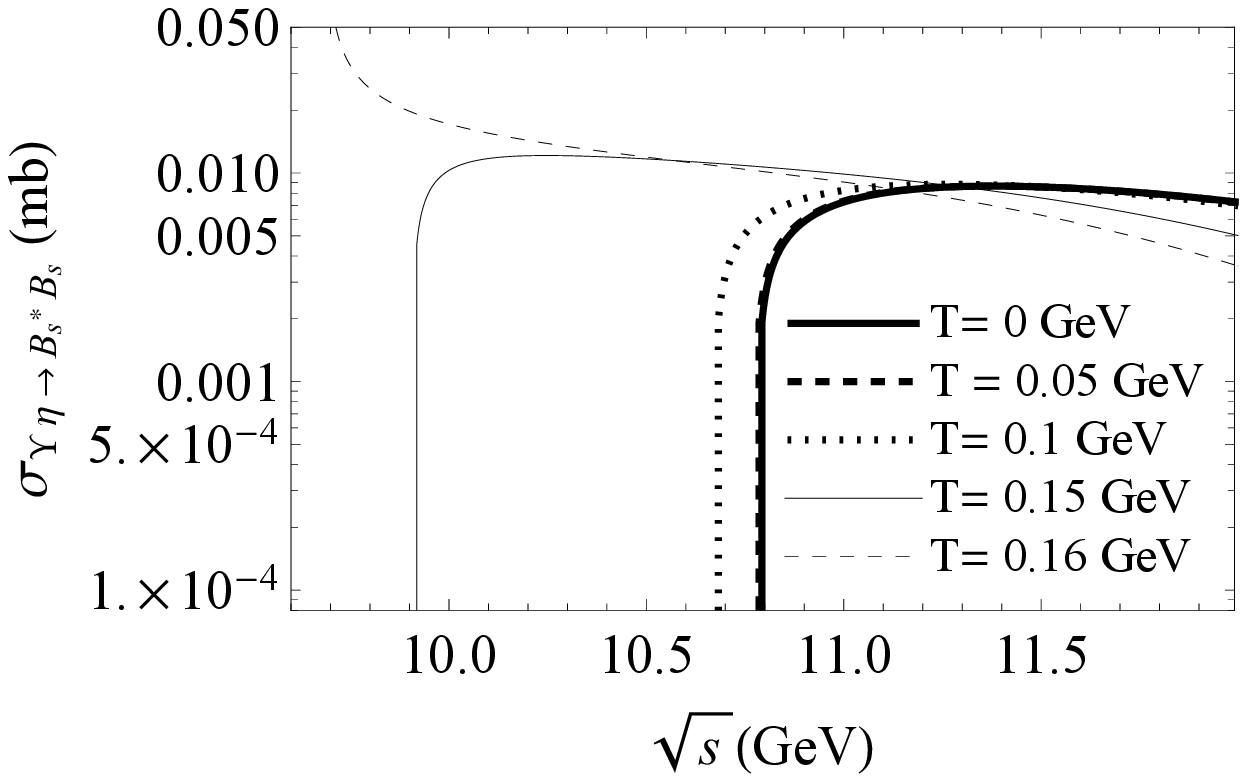}
	\\
	\includegraphics[width=5.9cm]{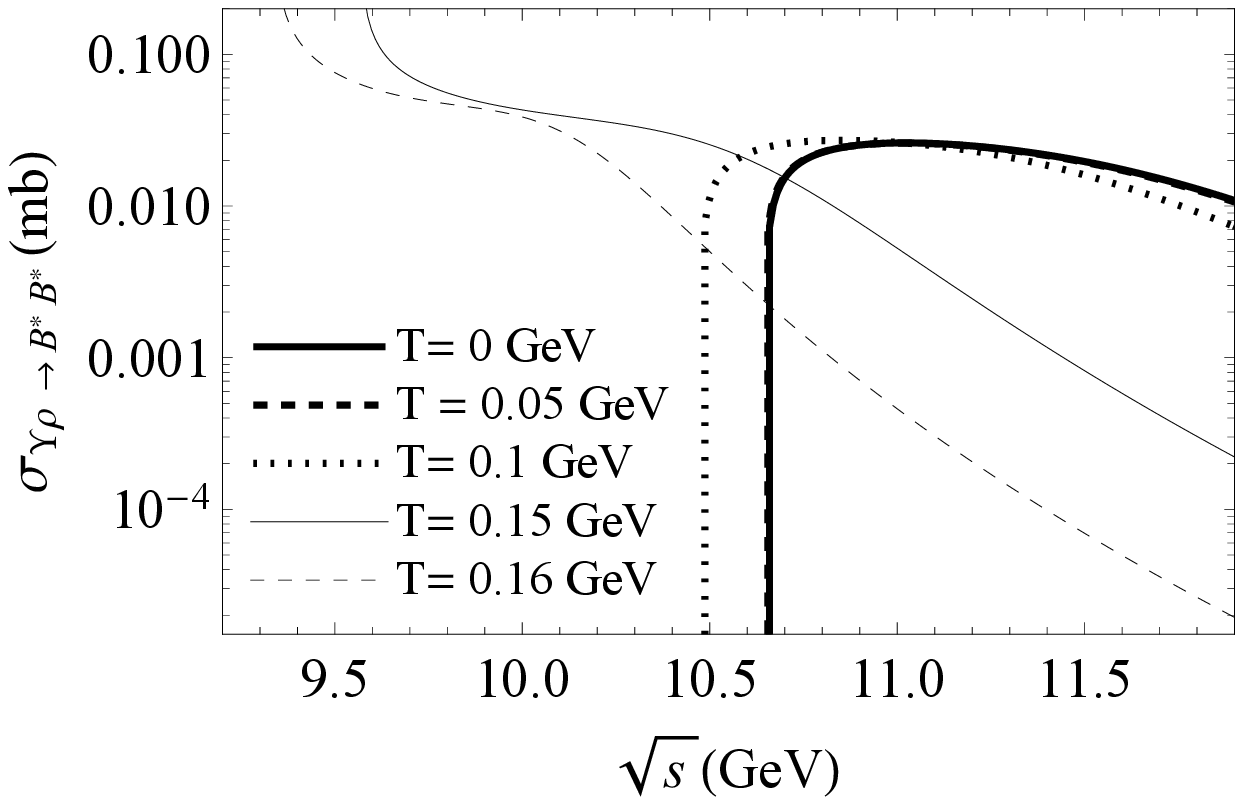}  
	\includegraphics[width=5.9cm]{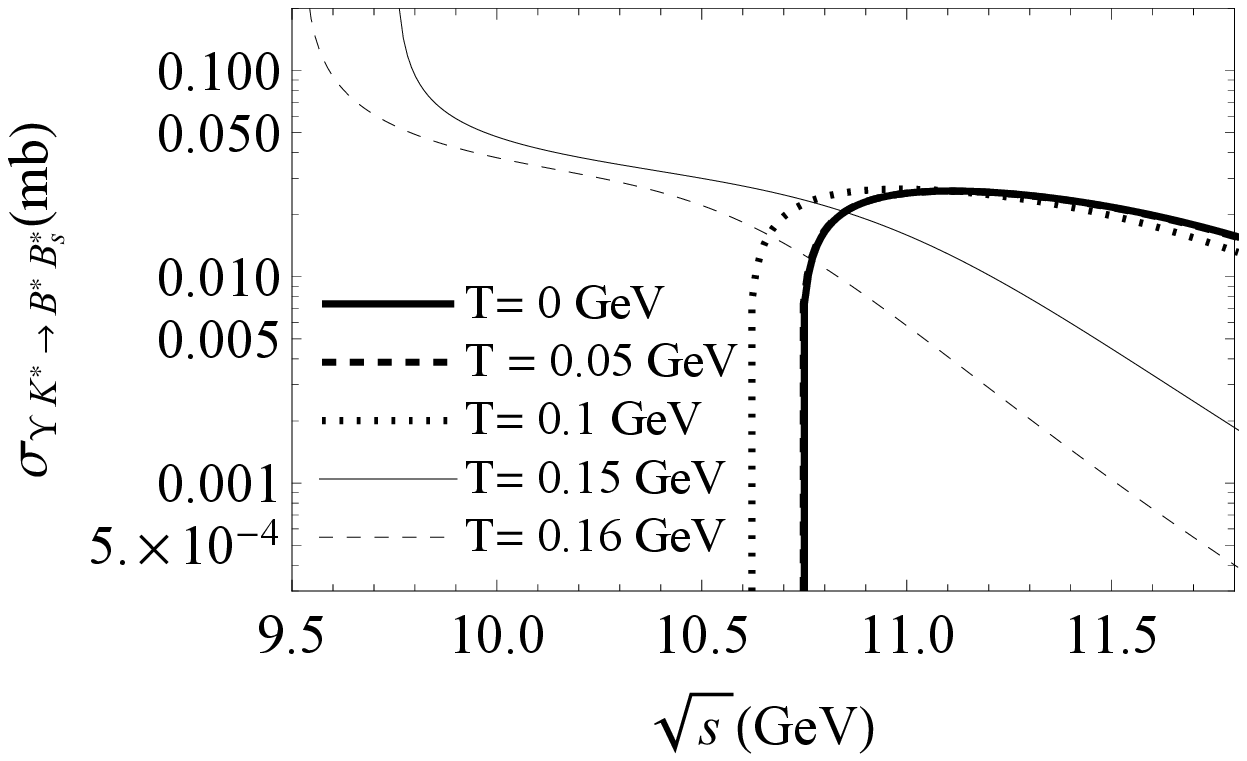} 
	\includegraphics[width=5.9cm]{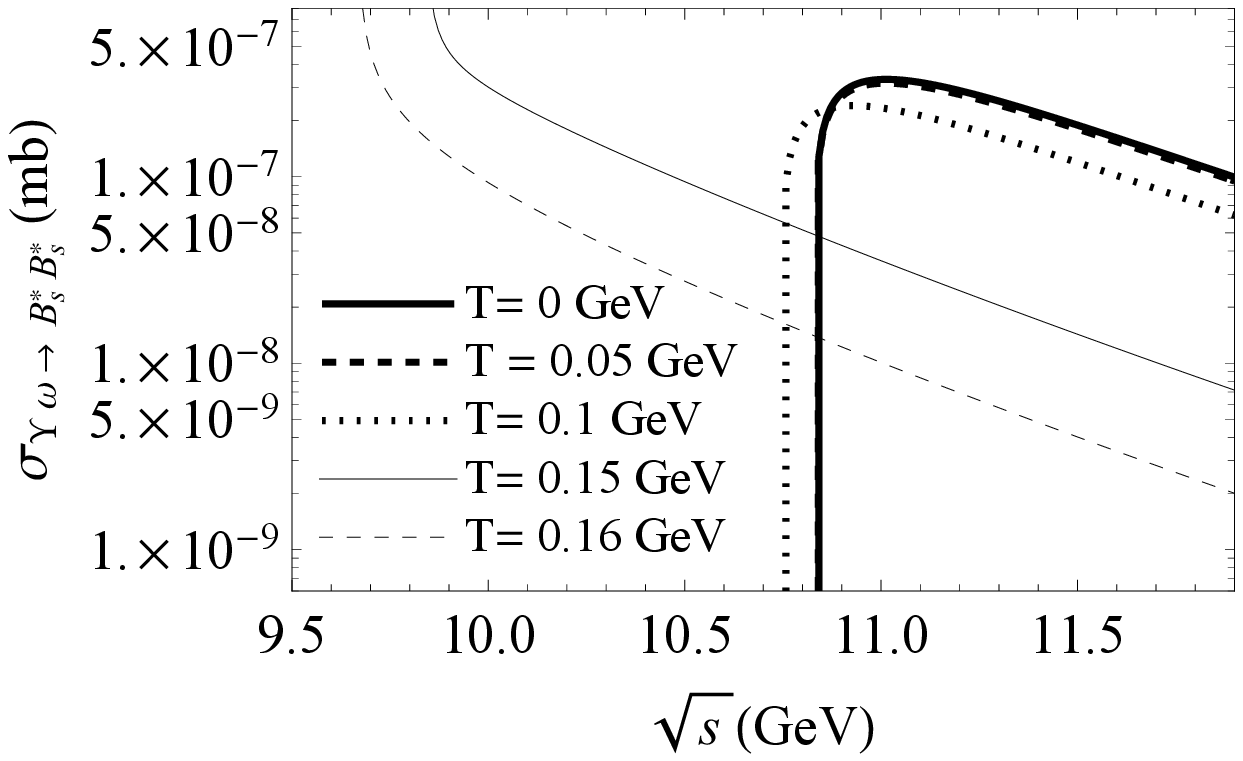}
	\caption{Unitarized cross sections for $\Upsilon X  (X=\pi,K,\eta,\rho,K^\ast,\omega)$  scatterings into relevant final states with open-charm mesons as a function of the CM energy $\sqrt{s}$, at different values of temperature. The curves corresponding to $T=0.05$ MeV almost coincide with those with $T=0$ MeV. }
	\label{fig:RelevanceYXFT}
\end{figure}

\end{widetext}

Coming back to the bottomonium dissociation reactions, the cross sections for the $\Upsilon $ absorption by light mesons into final states with open-bottom mesons at different temperatures are shown in Fig.~\ref{fig:RelevanceYXFT}. As in previous studied case of charmonium, the peak of the cross sections is driven to the left as the temperature increases,  due to the reduction of the threshold of the respective channel. Moreover,  we see that another effect takes place
in a more prominent way than in the case of charmonium (which is present only in the $J/\psi K \rightarrow D^* D_s$ reaction): some processes, which are endothermic at vanishing temperature, becomes exothermic with the elevation of the temperature, because the sum of the masses of particles in final state turns to be smaller than that in initial state. More explictly, for these mentioned reactions, like the $\Upsilon \pi \rightarrow (B \bar B^\ast + c.c.)$ (endothermic at $T=0$), the cross sections have a peak near the threshold energy, but at higher temperatures they come to be infinite at the threshold. 

We finish this section by discussing some points. The first one is that the interactions in the quarkonia absorption reactions are obviously dependent on the effective formalism considered, which determines the magnitudes and energy dependences of the cross sections. A comparison of our findings with those from previous works based in different approaches has been useful to quantify this issue. 
In addition, the validity of the present approach deserves to be highlighted: it seems reasonable in the low-energy range, in view of the use of $s$-wave projected amplitudes with lowest-order extended chiral Lagrangians. 
Furthermore, the thermal effects are also influenced by the finite-temperature framework as well as the parametrization of the thermal masses employed. 
In this sense, it is worthy noting that we have not taken into consideration in the unitarization procedure  the width $\Gamma$ of the particles. Since $\Gamma$ can suffer an increase at higher temperatures~\cite{Cleven:2017fun,Montana:2020lfi}, this might yield important modifications in the cross sections. 
Notwithstanding these points, our results have provided estimations of the most relevant interactions between the $\Upsilon$ and the hadronic medium composed of the lightest mesons, and also suggested that the modifications in the cross sections for the quarkonia absorption by light mesons due to thermal medium in some reactions are not negligible and should not be underestimated in physical situations like heavy-ion collisions.

%
%
%
%
%
%
%
%

\section{Concluding Remarks}
\label{Conclusions} 

In this work we have continued a previous study performed in the charmonium sector by evaluating the interactions of bottomonium with surrounding hadronic medium. The cross-sections for $\Upsilon $ absorption by light mesons ($\pi, K, \eta, \rho, K^\ast, \omega$) have been evaluated using unitarized coupled channel amplitudes, and a comparison of our results with existing works employing different approaches has been performed. 

The unitarized coupled channel amplitudes with $s$-wave projected kernels acquire non-vanishing cross sections via the meson loops, with a peak just after the threshold and a controlled decrease as energy grows. For the investigated cases, the processes with bottomed mesons or a bottomonium in the final state are the most relevant contributions, whereas the other contributions have negligible magnitudes and are strongly suppressed. 
Besides, in general the scatterings $\Upsilon P \rightarrow~\text{All}$ and $\Upsilon V \rightarrow \text{All}$ have comparable magnitudes in the considered range of center-of-mass energy.
With respect to the preceding works~\cite{Abreu:2018mji,Lin:2000ke}, which have analyzed only the processes involving $\Upsilon $ absorption by $ \pi, \rho$ mesons making use of distinct form-factors and cutoffs, our findings predict smaller magnitudes for these specific cross sections. 

Modifications due to thermal effects have also discussed using a simplified version of ITF and thermal masses. For completeness both charmonium and bottomonium sectors have been examined, 
and as the temperature increases, the position and magnitude of the peak of the cross sections are affected. This suggests that the thermal effects might be relevant in quarkonia absorption processes under certain situations, like in the heavy-ion collision environment. 

%

\begin{acknowledgements}

We are grateful to E. Cavalcanti for support and discussions. The authors would like to thank the Brazilian funding agencies CNPq (Contracts
No. 308088/2017-4 and No. 400546/2016-7) and FAPESB (Contract No. INT0007/2016) for financial support.
\end{acknowledgements} 


\begin{thebibliography}{99}%

\bibitem{Matsui:1986dk}  T. Matsui and H. Satz, Physics Letters B 178 (1986) 416.

\bibitem{RAPP2010209} R. Rapp, D. Blaschke, and P. Crochet, Progress in Particle and Nuclear Physics 65 (2010) 209.

\bibitem{Braun-Munzinger}P. Braun-Munzinger, V. Koch, T. Schafer, and J. Stachel, Physics Reports 621 (2016) 76.

\bibitem{Adams:2005dq}
J.~Adams \textit{et al.} [STAR],
Nucl. Phys. A \textbf{757} (2005), 102-183
doi:10.1016/j.nuclphysa.2005.03.085
[arXiv:nucl-ex/0501009 [nucl-ex]].

\bibitem{Alessandro:2004ap}
B.~Alessandro \textit{et al.} [NA50],
Eur. Phys. J. C \textbf{39} (2005), 335-345
doi:10.1140/epjc/s2004-02107-9
[arXiv:hep-ex/0412036 [hep-ex]].


                                             
\bibitem{Abelev:2013ila} B. B. Abelev {\it et al.} (ALICE Collaboration), Phys. Lett. B734 (2014) 314.

\bibitem{Adam:2016rdg} J. Adam {\it et al.} (ALICE Collaboration), Phys. Lett. B766 (2017) 212.


\bibitem{Chatrchyan:2012lxa}
S.~Chatrchyan \textit{et al.} [CMS],
Phys. Rev. Lett. \textbf{109} (2012), 222301
[erratum: Phys. Rev. Lett. \textbf{120} (2018) no.19, 199903]
doi:10.1103/PhysRevLett.109.222301
[arXiv:1208.2826 [nucl-ex]].

\bibitem{Zha:2017xsm} W. Zha and Z. Tang, Nucl. Part. Phys. Proc. 289 (2017) 83.


\bibitem{Wong:1999zb} C.-Y. Wong, E. S. Swanson, and T. Barnes, Phys. Rev. C62 (2000) 045201.
  
\bibitem{Wong:2001td} C.-Y. Wong, E. S. Swanson, and T. Barnes, Phys. Rev. C65 (2001) 014903; Erratum: Phys. Rev.C66 (2002) 029901.


\bibitem{PhysRevC.58.2994} S. G. Matinyan and B. Muller, Phys. Rev. C 58 (1998) 2994.


\bibitem{PhysRevC.61.031902} K. L. Haglin, Phys. Rev. C 61 (2000) 031902.
 
 
 \bibitem{Braun-Munzinger2000} P. Braun-Munzinger and K. Redlich, Eur. Phys. J. C 16  (2000) 519.
 
 
\bibitem{PhysRevC.62.034903} Z. Lin and C. M. Ko, Phys. Rev. C 62  (2000) 034903.



\bibitem{PhysRevC.63.065201} K. L. Haglin and C. Gale, Phys. Rev. C 63 
(2001) 065201.
  

\bibitem{PhysRevC.63.034901} Y. Oh, T. Song, and S. H. Lee, Phys. Rev. C 63
(2001) 034901.


\bibitem{PhysRevC.68.014903} T. Barnes, E. S. Swanson, C.-Y. Wong, and X.-M. Xu,
Phys. Rev. C 68 (2003) 014903.
 
 
\bibitem{Oh:2002vg} Y.-s. Oh, T.-s. Song, S. H. Lee, and C.-Y. Wong, J. Korean Phys. Soc. 43 (2003) 1003.
  
  


\bibitem{PhysRevC.68.035208} F. O. Duraes, H. Kim, S. H. Lee, F. S. Navarra, and M. Nielsen, Phys. Rev. C 68 (2003) 035208.



\bibitem{Maiani:2004py} L. Maiani, F. Piccinini, A. D. Polosa, and V. Riquer, Nucl. Phys. A741 (2004) 273.
 
 
\bibitem{Maiani:2004qj} L. Maiani, F. Piccinini, A. D. Polosa, and V. Riquer,
Nucl. Phys. A748 (2005) 209. 

\bibitem{DURAES200397} F. O. Dures, S. H. Lee, F. S. Navarra, and M. Nielsen,
Phys. Lett. B 564 (2003) 97.


\bibitem{PhysRevC.70.055203} A. Bourque, C. Gale, and K. L. Haglin, Phys. Rev. C
70 (2004) 055203.


\bibitem{PhysRevC.72.024902} F. Carvalho, F. O. Duraes, F. S. Navarra, and
M. Nielsen, Phys. Rev. C 72 (2005) 024902. 


\bibitem{PhysRevD.72.034002}T. Song and S. H. Lee, Phys. Rev. D 72 (2005) 034002.

\bibitem{Capella2008} A. Capella, L. Bravina, E. G. Ferreiro, A. B. Kaidalov,
K. Tywoniuk, and E. Zabrodin, Eur. Phys. J. C58 (2008) 437.

\bibitem{CASSING20011} W. Cassing, L. Kondratyuk, G. Lykasov, and M. Rz-
janin, Phys. Lett. B 513 (2001) 1.


\bibitem{doi:10.1142/S0218301308010507} O. Linnyk, E. L. Bratkovskaya, and W. Cassing, Int. J. Mod. Phys. E 17 (2008) 1367.
 
 
 \bibitem{PhysRevC.85.064904} J. Zhou and X.-M. Xu, Phys. Rev. C 85 (2012) 064904.
  
  
  
 \bibitem{MITRA201675} S. Mitra, S. Ghosh, S. K. Das, S. Sarkar, and J. e Alam,
Nucl. Phys. A 951 (2016) 75.
  
\bibitem{Liu2016} F.-R. Liu, S.-T. Ji, and X.-M. Xu, Journal of the Korean
Physical Society 69 (2016) 472.

\bibitem{PhysRevC.97.044902} L. M. Abreu, K. P. Khemchandani, A. M. Torres, F. S.
Navarra, and M. Nielsen, Phys. Rev. C 97 (2018) 044902.

\bibitem{PhysRevC.96.045201} M. Cleven, V. K. Magas, and A. Ramos, Phys. Rev. C
96 (2017) 045201.

\bibitem{Abreu:2018mji}
L.~M.~Abreu, E.~Cavalcanti and A.~P.~C.~Malbouisson,
Nucl. Phys. A \textbf{978}, 107-122 (2018)
doi:10.1016/j.nuclphysa.2018.08.001
[arXiv:1808.02115 [hep-ph]].


\bibitem{Lin:2000ke}
Z.~w.~Lin and C.~M.~Ko,
Phys. Lett. B \textbf{503}, 104-112 (2001)
doi:10.1016/S0370-2693(01)00092-2
[arXiv:nucl-th/0007027 [nucl-th]].

\bibitem{Abreu:2018mnc}
L.~M.~Abreu, F.~S.~Navarra and M.~Nielsen,
Phys. Rev. C \textbf{101} (2020) no.1, 014906
doi:10.1103/PhysRevC.101.014906
[arXiv:1807.05081 [nucl-th]].

\bibitem{Roca:2005nm} L. Roca, E. Oset, and J. Singh, Phys. Rev. D72 (2005) 014002.

\bibitem{Gamermann:2006nm}
D.~Gamermann, E.~Oset, D.~Strottman and M.~J.~Vicente Vacas,
Phys. Rev. D \textbf{76}, 074016 (2007)
doi:10.1103/PhysRevD.76.074016
[arXiv:hep-ph/0612179 [hep-ph]].

\bibitem{Gamermann:2007fi} D. Gamermann and E. Oset, Eur. Phys. J. A33
(2007) 119. 


\bibitem{Dias:2014pva}
J.~M.~Dias, F.~Aceti and E.~Oset,
Phys. Rev. D \textbf{91}, no.7, 076001 (2015)
doi:10.1103/PhysRevD.91.076001
[arXiv:1410.1785 [hep-ph]].



\bibitem{Molina:2008jw}
R.~Molina, D.~Nicmorus and E.~Oset,
Phys. Rev. D \textbf{78}, 114018 (2008)
doi:10.1103/PhysRevD.78.114018
[arXiv:0809.2233 [hep-ph]].


\bibitem{Abreu2011} L. M. Abreu, D. Cabrera, F. J. Llanes-Estrada, and
J. M. Torres-Rincon, Ann. Phys. (NY) 326, (2011) 2737.

\bibitem{Abreu2013a} L. M. Abreu, D. Cabrera, and J. M. Torres-Rincon,
Phys. Rev. D 87 (2013) 034019.
 
 
\bibitem{Gao:2019idb}
R.~Gao, Z.~H.~Guo and J.~Y.~Pang,
Phys. Rev. D \textbf{100}, no.11, 114028 (2019)
doi:10.1103/PhysRevD.100.114028
[arXiv:1907.01787 [hep-ph]].


 
 \bibitem{Weinstein:1990gu} J. Weinstein and N. Isgur, Phys. Rev. D41
(1990) 2236.
 
 
\bibitem{Janssen:1994wn} G. Janssen, B. C. Pearce, K. Holinde, and J. Speth,
Physical Review D 52 (1995) 2690.


\bibitem{Oller:1997ti} J. Oller and E. Oset, Nucl. Phys. A 620 (1997) 438.


\bibitem{Oller:1998hw} J. A. Oller, E. Oset, and J. R. Pelaez, Phys. Rev. D59 (1999) 074001.

\bibitem{Bazavov:2017lyh}
A.~Bazavov, C.~Bernard, N.~Brown, C.~Detar, A.~X.~El-Khadra, E.~G\'amiz, S.~Gottlieb, U.~M.~Heller, J.~Komijani and A.~S.~Kronfeld, \textit{et al.}
Phys. Rev. D \textbf{98}, no.7, 074512 (2018)
doi:10.1103/PhysRevD.98.074512
[arXiv:1712.09262 [hep-lat]].

\bibitem{Aoki:2016frl}
S.~Aoki, Y.~Aoki, D.~Becirevic, C.~Bernard, T.~Blum, G.~Colangelo, M.~Della Morte, P.~Dimopoulos, S.~D\"urr and H.~Fukaya, \textit{et al.}
Eur. Phys. J. C \textbf{77}, no.2, 112 (2017)
doi:10.1140/epjc/s10052-016-4509-7
[arXiv:1607.00299 [hep-lat]].

\bibitem{Abreu:2017nuc}
L.~M.~Abreu, K.~P.~Khemchandani, A.~Mart\'\i{}nez Torres, F.~S.~Navarra, M.~Nielsen and A.~L.~Vasconcellos,
Phys. Rev. D \textbf{95}, no.9, 096002 (2017)
doi:10.1103/PhysRevD.95.096002
[arXiv:1704.08781 [hep-ph]].

\bibitem{Zhou:2012vv}
J.~Zhou and X.~M.~Xu,
Phys. Rev. C \textbf{85}, 064904 (2012)
doi:10.1103/PhysRevC.85.064904
[arXiv:1206.2440 [hep-ph]].
\bibitem{Ji:2015fva}
S.~T.~Ji, Z.~Y.~Shen and X.~M.~Xu,
J. Phys. G \textbf{42}, no.9, 095110 (2015)
doi:10.1088/0954-3899/42/9/095110
[arXiv:1507.04262 [hep-ph]].


\bibitem{Cleven:2017fun}
M.~Cleven, V.~K.~Magas and A.~Ramos,
Phys. Rev. C \textbf{96} 045201 (2017)
doi:10.1103/PhysRevC.96.045201.



\bibitem{Gu:2018swy}
X.~W.~Gu, C.~G.~Duan and Z.~H.~Guo,
Phys. Rev. D \textbf{98}, no.3, 034007 (2018)
doi:10.1103/PhysRevD.98.034007
[arXiv:1803.07284 [hep-ph]].

\bibitem{Montana:2020lfi}
G.~Monta\~na {\it et al.}
Phys. Lett. B \textbf{806}, 135464 (2020)
doi:10.1016/j.physletb.2020.135464.


\bibitem{LeBellac} M. Le Bellac, Thermal Field Theory (Cambridge University Press, Cambridge, England, 1996).

\bibitem{Schenk:1993ru}
A.Schenk,
Phys. Rev. D \textbf{47} (1993), 5138
doi: 10.1103/PhysRevD.47.5138. 

\bibitem{Nicola:2014eda}
A.~G\'omez Nicola and R.~Torres Andr\'es,
Phys. Rev. D \textbf{89} (2014) no.11, 116009
doi:10.1103/PhysRevD.89.116009
[arXiv:1404.2746 [hep-ph]].

\bibitem{Cheng:2010fe}
M.~Cheng, S.~Datta, A.~Francis, J.~van der Heide, C.~Jung, O.~Kaczmarek, F.~Karsch, E.~Laermann, R.~D.~Mawhinney and C.~Miao, \textit{et al.}
Eur. Phys. J. C \textbf{71} (2011), 1564
doi:10.1140/epjc/s10052-011-1564-y
[arXiv:1010.1216 [hep-lat]].



\end{thebibliography}
%
%

\appendix*

%
%
%
%
%
%
%
%
%
%
%
%
%
%
%
%
%
%


%

\end{document}